\documentclass[12pt]{iopart}
\usepackage{iopams}
\usepackage{amssymb}
\usepackage{portland}
\usepackage{subfigure}
\usepackage{graphicx}
\usepackage{dcolumn}
\usepackage{epsfig}
\usepackage{amsfonts}
\usepackage{graphicx}
\usepackage{amssymb}

\begin{document}
\title{Causal Nature and Dynamics of Trapping Horizons in Black Hole Collapse}
\author{Alexis Helou${}^{1,2}$, Ilia Musco${}^{3}$ and John C. Miller${}^{4}$} 


\address{ ${}^1$ Arnold Sommerfeld Center, Ludwig-Maximilians-Universit\"{a}t, 
Theresienstr. 37, 80333 M\"{u}nchen, Germany \\ 
${}^2$ AstroParticule et Cosmologie, Universit\'{e} Paris Diderot, CNRS, 
CEA, Observatoire de Paris, Sorbonne Paris Cit\'{e}, B\^{a}t. Condorcet, 10 
rue Alice Domon et L\'{e}onie Duquet, F-75205 Paris Cedex 13, France \\
${}^3$ Laboratoire Univers et Th\'{e}ories, UMR 8102 CNRS, Observatoire de 
Paris, Universit\'{e} Paris Diderot, 5 Place Jules Janssen, F-92190 Meudon, 
France \\
${}^4$ Department of Physics (Astrophysics), University of Oxford, Keble 
Road, Oxford OX1 3RH, UK \\ }
\vspace{0.25cm}
\address{E-mail: ilia.musco@obspm.fr}

\begin{abstract}
 In calculations of gravitational collapse to form black holes, trapping 
horizons (foliated by marginally trapped surfaces) make their first 
appearance either within the collapsing matter or where it joins on to a 
vacuum exterior. Those which then move outwards with respect to the 
matter have been proposed for use in defining black holes, replacing the 
global concept of an ``event horizon'' which has some serious drawbacks 
for practical applications. We here present results from a study of the 
properties of both outgoing and ingoing trapping horizons, assuming 
strict spherical symmetry throughout. We have investigated their causal 
nature (i.e.\ whether they are spacelike, timelike or null), making 
contact with the Misner-Sharp-Hernandez formalism, which has often 
been used for numerical calculations of spherical collapse. We follow 
two different approaches, one using a geometrical quantity related to 
expansions of null geodesic congruences, and the other using the horizon 
velocity measured with respect to the collapsing matter. After an introduction 
to these concepts, we then implement them within numerical simulations 
of stellar collapse, revisiting pioneering calculations from the 1960s where 
some features of the emergence and subsequent behaviour of trapping 
horizons could already be seen. Our presentation here is aimed firmly at 
``real world'' applications of interest to astrophysicists and includes the 
effects of pressure, which may be important for the asymptotic behaviour 
of the ingoing horizon.
 \end{abstract}
 
Published in {\CQG}{\bf 34}, 135012 (2017)

\maketitle

\section{Introduction}
 \label{section_intro}

The usual concept of a black hole is associated with a region of 
spacetime from which nothing can ever escape, including light rays. Any 
observer who remains outside its outer boundary (called the \emph{event 
horizon}) can never know anything about what happens inside. In the 
simplest mathematical picture, black holes are envisaged as eternal 
unchanging vacuum objects for which one knows the entire global structure 
of the spacetime, but real black holes are not eternal and unchanging: 
they are born, interact with their surroundings and will probably 
eventually evaporate away.

Astrophysical black holes can be formed in various ways, with their masses 
spanning a large range, from very small ones formed in the early Universe 
up to supermassive ones formed in the centres of galaxies. The simplest 
formation scenario concerns collapse of a single object, such as a star or 
gas cloud, whose internal pressure becomes inadequate to support it. The 
collapse is a dynamical process and subsequent accretion or interactions 
with surrounding objects can lead to further dynamical changes. For 
calculating the evolution of dynamical black holes without having to know 
the entire global structure of the four-dimensional spacetime, they need to 
be characterized in terms of some quasi-local concept rather than by the 
global concept of an event horizon. The notion of trapped surfaces, from 
which null-rays cannot expand outwards, may provide such a quasi-local 
characterization, with the black hole being thought of as a region of 
closed trapped surfaces. (Note that trapped surfaces are ``quasi-local'' 
rather than local, both because they are extended rather than point-like, 
and also because it is necessary to move infinitessimally away from the 
surface in order to measure any expansion of null rays 
\cite{Visser:2014zqa,Krishnan:2013saa,Faraoni:2015ula}.) The issue of 
correctly defining the boundary of the trapped surfaces, or ``trapping 
boundary'', is an important and delicate one 
\cite{Hayward:1993wb,Booth:2005qc,Bengtsson:2010tj,Eardley:1997hk,BenDov:2006vw}, 
but it is not one which we are addressing in the present paper. Indeed, we 
are here focusing entirely on spherical symmetry, both as regards the 
spacetime and the trapped surfaces. We are not treating the so-called 
``non-round spheres'' (surfaces having spherical topology but not spherical 
symmetry) which are often discussed in connection with precisely defining 
the trapping boundary, even within spherically-symmetric spacetimes 
\cite{Wald_Iyer,Schnetter:2005ea}. We chose to restrict our attention to 
analyzing the behaviour of the spherical marginally-trapped surfaces 
appearing during collapse to form black holes in spherical symmetry, 
following common practice in the literature 
\cite{Dafermos:2004wr,Williams:2007tp,Booth:2005ng,Faraoni:2016xgy}.

There are several different quasi-local definitions of black hole 
horizons in the literature, e.g.\ the apparent horizon 
\cite{Hawking:1973uf}, the trapping horizon \cite{Hayward:1993wb}, the 
isolated horizon \cite{Ashtekar:2000sz}, the dynamical horizon 
\cite{Ashtekar:2002ag, Ashtekar:2003hk}, and the slowly-evolving horizon 
\cite{Booth:2003ji}. We recommend \cite{Ashtekar:2004cn} and 
\cite{Faraoni:2013aba} for extensive reviews of these concepts.

We recall that while physical objects are constrained to follow either 
timelike trajectories (if they have non-zero rest mass) or null 
trajectories (if they are massless), horizons do not have this constraint 
and could also be spacelike (i.e.\ they could be superluminal, moving 
outside the local lightcone). An alternative way of saying this is that 
the \emph{signature} of the horizon could be \emph{spacelike}, as well as 
\emph{null} or \emph{timelike}. A main focus of the work presented here 
is on following this causal nature of horizons during gravitational 
collapse to form a black hole. Within the different definitions of 
quasi-local horizons, the isolated horizon is always null, the dynamical 
horizon can only be spacelike or null and the slowly evolving horizon has 
to remain close to being null. These different definitions each refer to 
restricted classes, and we preferred to work here with the more general 
concept of \emph{trapping horizons}, as defined by Hayward 
\cite{Hayward:1993wb}, which can have any signature. As we will see, this 
notion is very general and applies also in cosmology and for other 
situations where trapping occurs \cite{Hayward:1993wb, Helou:2015zma}. 
The full trapping horizon is a 3D surface in 4D spacetime, but in order 
to follow collapse as a Cauchy problem, with the specification of initial 
data which is then evolved forward in time, we follow the standard 
practice of considering the horizon as a 2D surface which evolves with 
time. (Whenever we are discussing simulation results here, we will 
always be using the term ``horizon'' in this sense.) This 2D surface is 
usually called an ``apparent horizon'', but in the literature the terms 
apparent/trapping horizons are often used as synonymous \cite{Visser:2014zqa,
Poisson}.

As a preliminary for our analysis, we relate the standard geometrical 
machinery used for studying trapping horizons with the 
Misner-Sharp-Hernandez hydrodynamical formalism 
\cite{Misner:1964je,Misner:1966hm,May:1966zz}, which we use later for 
studying dynamical black holes in spherical symmetry\footnote{This 
formalism (using a diagonal metric) was first presented by Misner \& 
Sharp \cite{Misner:1964je} and then re-expressed with extended 
terminology in a paper by Hernandez \& Misner \cite{Misner:1966hm} (where 
they also presented an alternative approach using an ``observer time'' 
null slicing). Another diagonal-metric formulation was developed 
separately by May \& White \cite{May:1966zz} who then implemented it in 
their numerical investigations. The extended Misner-Sharp and May \& 
White formulations are essentially equivalent but the notations are 
different in some respects. What we describe here as the 
``Misner-Sharp-Hernandez'' approach is, in fact, a composite of these 
two.}. We show that the geometrical and Misner-Sharp-Hernandez 
approaches are completely consistent with each other but give alternative 
frameworks for understanding the behaviour being studied. Following either 
approach, one reaches the result that the condition $R=2M$ (where $M$ is 
the ``mass contained within radius $R$'', see Section 
\ref{section_trapping_horizon} for more precise definitions) is not only 
associated with the event horizon of the vacuum Schwarzschild 
metric\footnote{In stationary spacetimes, there is no difference between 
event and quasi-local horizons, and the characteristic $R=2M$ is usually 
attributed to the event horizon for historical precedence. But in 
dynamical cases where the two horizons are different, $R=2M$ holds for 
the quasi-local horizon.}, but applies for \emph{all} spherically 
symmetric trapping horizons in spherically symmetric spacetimes, whether 
of dynamical black holes or in cosmology. In collapse to form black 
holes, the horizons are usually seen to form in pairs: the two horizons 
emerge from a single marginally trapped surface, with one of them then 
moving inwards and the other moving outwards (with respect to the 
matter). The outward-moving one is often the only object of study, under 
the names ``dynamical'' or ``apparent'' horizon (further considerations 
apply here; see \cite{Ashtekar:2004cn}). However, the ingoing one can 
also be of interest (cf. \cite{Jaramillo:2012}).

When following the geometrical and Misner-Sharp-Hernandez 
approaches, we use two different (but related) quantities for discussing 
the causal nature of the trapping horizons: with the geometrical approach 
we use the quantity $\alpha$, as in \cite{Dreyer:2002mx}, while with the 
Misner-Sharp-Hernandez approach we use the horizon three velocity 
$v_H$, measured in the local comoving frame of the fluid. By tracking the 
$R=2M$ condition, we have followed the evolution of $\alpha$ and $v_H$ 
for each horizon during numerical simulations for collapse of idealized 
stellar models, observing how varying the initial density profiles and 
the equation of state affects the horizon evolution and its signature. 
Our results supplement ones from previous work studying pressureless 
fluid collapse \cite{Booth:2005ng} and a case including angular momentum 
\cite{Schnetter:2006yt}. As well as presenting results which highlight 
some important new features and cast light on aspects of the previous 
literature, our work also contributes new insights gained by using the 
parallel geometrical and Misner-Sharp-Hernandez approaches.

The structure of the paper is as follows: in Section 
\ref{section_trapping_horizon}, we first review the fundamentals of the 
Misner-Sharp-Hernandez approach and the concept of trapped 
surfaces, and then demonstrate the connection between the geometrical and 
Misner-Sharp-Hernandez formalisms. In Section \ref{section_causal}, 
we introduce the definition of the parameters $\alpha$ and $v_H$, 
explaining how they are related to each other. In Sections 
\ref{section_black_hole} and \ref{General_perspective}, which contain the 
main results of the paper, we implement these concepts within numerical 
simulations of stellar collapse, studying the behaviour of the ingoing and 
outgoing horizons in various cases, and then give a broader perspective 
using a hyperbola diagram. Finally, Section \ref{section_conclusion} 
contains a summary and conclusions. Throughout, we work entirely within 
spherical symmetry for both the spacetime and the trapped surfaces, and 
we use the standard convention of setting $c=G=1$.

\section{Trapping Horizons in the Misner-Sharp-Hernandez approach}
 \label{section_trapping_horizon}
 
\subsection{Introduction to the Misner-Sharp-Hernandez formalism}
The Schwarzschild metric, usually written as 
 \begin{equation} ds^2 = - \left(1-\frac{2M}{R}\right) dt^2 + 
\left(1-\frac{2M}{R}\right)^{-1} dR^2 + R^2d\Omega^2 \ , 
\label{Schw-metric} 
\end{equation}
 where $d\Omega^2=d\theta^2 +\sin^2(\theta)d\varphi^2$ is the element of 
a 2-sphere, is the unique static, spherically symmetric solution of the 
vacuum Einstein equations describing the spacetime outside an object 
having mass $M$ but no charge or angular momentum. It may represent the 
spacetime outside an extended object (e.g.\ a star) but, if 
Eq.(\ref{Schw-metric}) is taken to hold \emph{everywhere}, then it 
represents a black hole with the mass $M$ being collapsed to $R=0$, 
creating a spacetime singularity there. If Eq.(\ref{Schw-metric}) is 
taken to hold for all time, the black hole is said to be \emph{eternal}, 
and the Schwarzschild radius $R_S = 2M$ then gives the location of its 
event horizon. For studying horizon evolution in dynamical situations, 
involving a fluid medium, we are following the Misner-Sharp-Hernandez approach 
\cite{Misner:1964je,Misner:1966hm,May:1966zz}, using a metric of the form
 \begin{equation}
ds^2 = - a^2(r,t) dt^2 + b^2(r,t) dr^2 + R^2(r,t)d\Omega^2 \ ,
\label{eq_metric_MS}
\end{equation}
 where the radial coordinate $r$ is taken to be comoving with the 
collapsing fluid, which then has four-velocity $u^a = (a^{-1},0,0,0)$, 
and $t$ is sometimes referred to as ``cosmic time''. This metric 
corresponds to an orthogonal comoving foliation of the spacetime, where 
$a$, $b$ and $R$ are positive definite functions of $r$ and $t$; 
$R$ is called the ``circumference coordinate'' in \cite{Misner:1964je} 
(being the proper circumference of a sphere with coordinate labels 
$(r,t)$, divided by $2\pi$ - this is equivalent to the quantity referred 
to as the ``areal'' coordinate), and $d\Omega$ is the element of a 
2-sphere of symmetry. The metric \ref{eq_metric_MS} can apply to any 
spherically symmetric spacetime; in the particular case of a homogeneous 
and isotropic universe, it can be rewritten in the form of the FLRW 
metric.

In the Misner-Sharp-Hernandez approach, two basic differential 
operators are introduced
 \begin{equation}
D_t \equiv \frac{1}{a} \frac{\partial}{\partial t}  \ \ 
\textrm{and} \ \ 
D_r \equiv \frac{1}{b} \frac{\partial}{\partial r}  \label{operators} \ ,
\end{equation}
 representing derivatives with respect to proper time and radial proper 
distance in the comoving frame of the fluid. These operators are then 
applied to $R$, and doing this gives the quantities
 
\begin{eqnarray}
 &U \equiv D_t R = \frac{1}{a} \frac{\partial R}{\partial t} \ ,  
\label{U_def} \\ 
 &\Gamma \equiv D_r R = \frac{1}{b} \frac{\partial R}{\partial r} 
\label{Gamma_def} \ ,
\end{eqnarray}
 with $U$ being the radial component of four-velocity in an ``Eulerian'' 
(non comoving) frame where $R$ is used as the radial coordinate, and 
$\Gamma$ being a generalized Lorentz factor (which reduces to the 
standard one in the special relativistic limit). These two quantities are 
related to the Misner-Sharp-Hernandez mass $M$ (mathematically 
appearing as a first integral of the $G^0_0$ and $G^1_1$ components of 
the Einstein equations) by the constraint equation
 \begin{equation}
 \Gamma^2 = 1 + U^2 - \frac{2M}{R} \ ,
 \label{eq_MS_mass}
\end{equation}
 where the interpretation of $M$ as a mass becomes transparent when the 
form of the stress energy tensor, on the right hand side of the Einstein 
equations, is specified. In this paper we will be considering matter 
described as a perfect fluid, with the stress energy tensor
 \begin{equation}
T^{ab} = (e+p)u^a u^b + pg^{a b} \,,
 \label{eq_stress}
\end{equation}
 where $e$ and $p$ are the fluid energy density and pressure, as measured 
in the comoving frame of the fluid, and $u^a$ is the fluid 
four-velocity; $M$ is then given by 
 \begin{equation}
M = \int_0^R 4\pi R^2e\,dR \ .
\end{equation}
It has been argued in \cite{Misner:1974qy} and \cite{Hayward:1994bu} 
that, in spherically symmetric spacetimes, this mass can be seen as a 
local gravitational energy for matter internal to a sphere of 
circumferential radius $R$. It can be written in a covariant way as
\begin{equation}
 M=\frac{R}{2} \left(1-\nabla^a R \,\nabla_a R \right)  \ ,
 \label{eq_MS_energy}
\end{equation}
with $\nabla^a R \,\nabla_a R=\Gamma^2-U^2$.

\subsection{Trapping Horizons}
 The definitions of the various quasi-local horizons mentioned in the 
Introduction are based on the concept of trapped surfaces 
\cite{Penrose:1964wq}, and on the limit notion of a marginally trapped 
surface. We first give a review of these ideas within the geometrical 
approach, where they have their origin, and then demonstrate how they 
transfer to the Misner-Sharp-Hernandez approach\footnote{In the 
original papers by Misner and his collaborators where trapped surfaces 
and ``observer time'' were discussed (see 
\cite{Misner:1966hm,Misner:1969br} and others), null rays were spoken of 
in terms of idealized light rays which are not affected by any matter 
through which they pass. This terminology was then followed by other 
authors using similar methodology. While this was useful for exposition, 
we will use only the term ``null ray'' here.}.

We follow here the geometrical approach and terminology of Hayward 
\cite{Hayward:1993wb}. Consider a spatial 2-sphere $S$ in any 
spacetime: there are two unique future null directions normal to $S$ used 
in the double-null formalism. One may then compute the expansion $\theta$ 
of a null geodesic congruence (i.e.\ the expansion/contraction of a bundle of 
null-rays) by Lie-dragging the surface $S$ along one of our two null directions 
(see \cite{Gourgoulhon:2008pu} for more details and a pictorial explanation). 
If this expansion $\theta$ is negative, then the area of the surface $S$ 
is shrinking in the chosen null direction, and the null-rays are 
converging. Instead if $\theta>0$, the area of $S$ is growing in the 
direction under consideration, and the null-rays are diverging. In the 
limiting case of $\theta=0$, the area of $S$ is not varying in the 
corresponding direction. Our common intuition from Minkowski spacetime is 
that one null direction will have a strictly positive $\theta$ while the 
other has a strictly negative $\theta$, corresponding respectively to the 
divergence of outgoing null-rays and the convergence of ingoing 
null-rays. In Minkowski spacetime we can have only this type of 
configuration, where the compact 2-surface $S$ is called an 
untrapped or ``normal'' surface.

However there are spacetimes which contain some compact 
2-surfaces for which both expansions have the same 
sign: when both are negative, the surface is 
said to be ``future-trapped'' (as occurs inside a black hole); when 
both are positive, the surface is said to be ``past-trapped'' (as 
occurs in an expanding universe). The transition from a normal region, 
where compact 2-surfaces are of the Minkowski type, to future-trapped or 
past-trapped regions is characterized by the change of sign of an 
expansion in one of the two null directions. Let us label this particular 
null direction by the letter $v$ (which stands for ``vanishing'') and the 
other radial null direction by $nv$ (which stands for ``non-vanishing''): 
$\theta_v$ can change sign and vanish while $\theta_{nv}$ can never 
change sign or vanish. The surfaces with $\theta_v=0$ are called 
``marginally trapped surfaces''. A quasi-local horizon is then a 
hypersurface foliated by marginally trapped surfaces.

\begin{figure}[t] 
\vspace{0.25cm}
 \centering  
 \includegraphics[width=8cm]{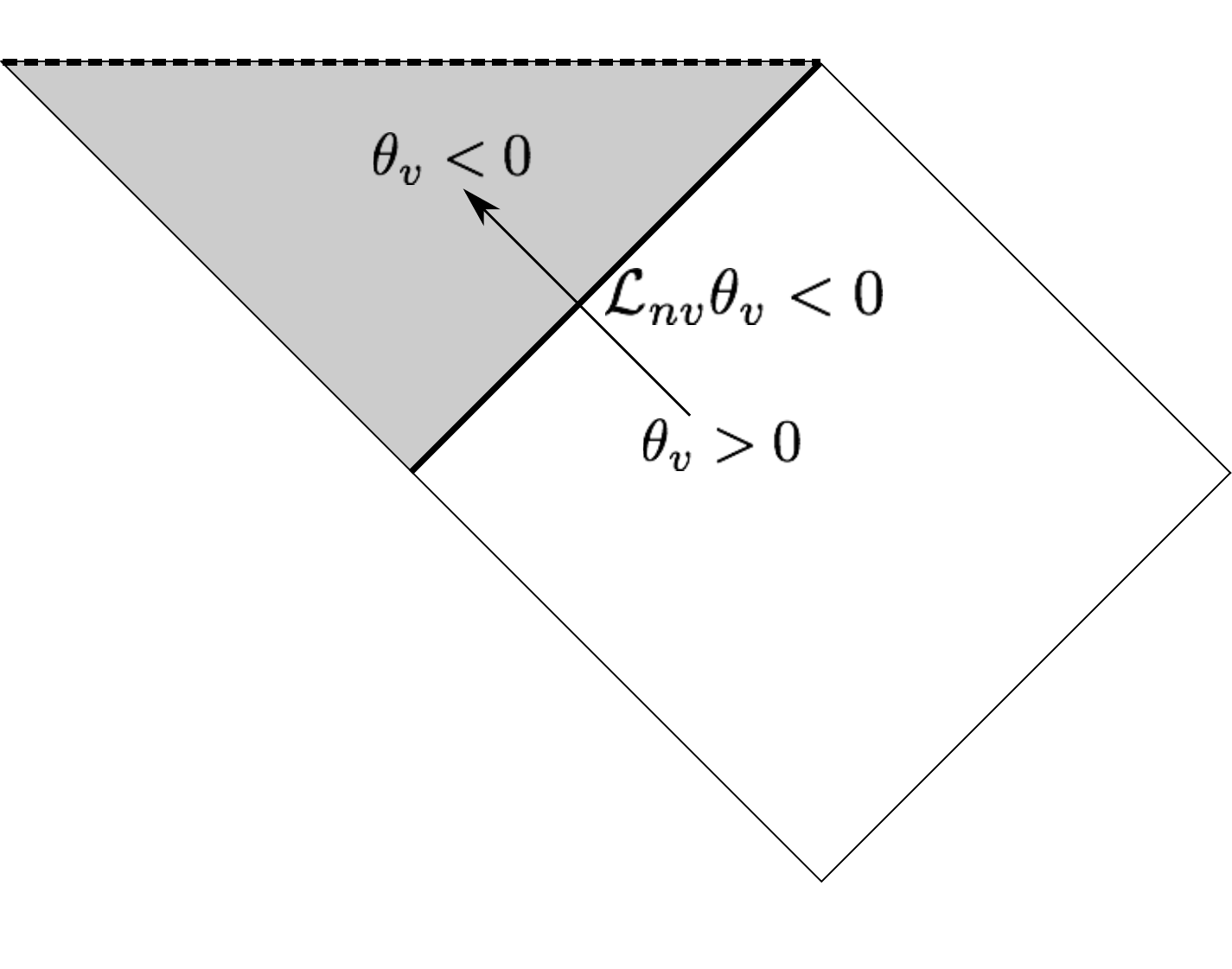}
 \vspace{-0.15cm}
   \caption{Penrose-Carter diagram for a Schwarzschild black hole. The 
   dashed line represents the $R=0$ singularity and the heavy 
   solid line represents a \emph{trapping horizon} that is \emph{future} 
   (\mbox{$\theta_{nv}=\theta_-<0$}) and \emph{outer} 
   ($\mathcal{L}_{nv}\theta_v<0$, represented here by the null arrow). 
   }
    \label{fig_Schwarzschild}
\end{figure}

As explained in the Introduction, among the family of quasi-local 
horizons, the \emph{trapping horizon} seems the best suited for the 
purposes of the present paper. Following the definition introduced by 
Hayward \cite{Hayward:1993wb}, a trapping horizon is the closure of a 
three-surface $H$ foliated by marginally trapped surfaces ($\theta_v=0$) 
on which $\theta_{nv} \neq 0$ and where the Lie derivative 
$\mathcal{L}_{nv}\theta_v \neq 0$ (see Figure \ref{fig_Schwarzschild} for 
a more intuitive understanding of the Lie derivative). A further 
characterisation given in \cite{Hayward:1993wb} discriminates among 
different cases by saying that the trapping horizon is:
 \begin{itemize}
 \item \emph{outer} if $\mathcal{L}_{nv}\theta_v < 0$.
 \item \emph{inner} if $\mathcal{L}_{nv}\theta_v > 0$.
 \item \emph{future} if $\theta_{nv} < 0$. Then the ``non-vanishing'' 
direction is the ingoing radial null direction and the ``vanishing'' 
direction is the outgoing radial null direction. Denoting the outgoing 
and ingoing directions with $+$ and $-$ indices respectively, we then 
have: $\theta_- < 0$ and $\theta_+ = 0$. This is the case for a black 
hole.
 \item \emph{past} if $\theta_{nv} > 0$. Then the ``non-vanishing'' 
direction is the outgoing null radial direction, i.e.\ $\theta_+ > 0$ and 
$\theta_- = 0$. This is the case for an expanding universe.
 \end{itemize}
 All of the expansions $\theta$ and Lie derivatives $\mathcal{L}$ used 
above are evaluated at the horizon. We will use this terminology in the 
remainder of the paper, and apply it to 3D hypersurfaces as well as 2D 
surfaces for convenience, as explained in the Introduction.

Using this formalism within the Misner-Sharp-Hernandez approach, we 
can compute the expansion along the two future radial null directions. 
For the metric (\ref{eq_metric_MS}), the future outgoing radial null vector 
is
 \begin{eqnarray}
 k^a= \left( \frac{1}{a},+\frac{1}{b},0,0 \right) \ , \nonumber 
\end{eqnarray}
and the future ingoing radial null vector is
\begin{eqnarray}
 l^a= \left( \frac{1}{a},-\frac{1}{b},0,0 \right) \ , \nonumber 
\end{eqnarray}
 with normalisation $k^a l_a = -2$. We can then define the induced metric 
on the 2-spheres of symmetry, denoted by $h_{ab}$, as
 \begin{equation}
 h_{ab} = g_{ab} + \frac{1}{2}(k_a l_b + l_a k_b) \ ,
 \label{eq_metric_2sphere}
\end{equation}
 and compute the expansions $\theta_\pm$ of outgoing and ingoing bundles 
of null-rays
 \begin{eqnarray}
 \theta_{+}&= h^{cd} \nabla_c k_d =\frac{2}{R} (U + \Gamma) \ ,  
  \label{eq_theta+}
\end{eqnarray}
and
\begin{eqnarray}
 \theta_{-}&= h^{cd} \nabla_c l_d =\frac{2}{R} (U - \Gamma) \ .
  \label{eq_theta-}
\end{eqnarray}
 For a \emph{future trapping horizon} (the black hole case), $\theta_+$ 
vanishes and $\Gamma=-U$ at the horizon, while for a \emph{past trapping 
horizon} (in an expanding universe) $\theta_-=0$ and $\Gamma=+U$ at the 
horizon. In both cases 
 \begin{equation}
  \theta_+\theta_- =\theta_v\theta_{nv} =\frac{4}{R^2} (U^2-\Gamma^2) \ ,
   \label{eq_theta-theta+}
\end{equation}
 with $\theta_v=0$ corresponding to $\Gamma^2 = U^2$ and so, using 
Eq.(\ref{eq_MS_mass}),  the result $R=2M$ (familiar for the event 
horizon of a Schwarzschild black hole) emerges as the condition for 
\emph{all} spherically symmetric trapping horizons in spherically 
symmetric spacetimes. The general nature of this condition was already 
noted in the 1960s by Misner and his colleagues 
\cite{Misner:1966hm,Misner:1969br} in the work which we are using as a 
basis for our present study.

An alternative way to arrive at this result is as follows. The general 
expression for changes in $R$ along a radial worldline
 \begin{equation}
 dR = \frac{\partial R}{\partial t}\,dt
    + \frac{\partial R}{\partial r}\,dr \,,
\label{RWL}
\end{equation}
 can be rewritten as
 \begin{equation}
 \frac{1}{a}\frac{dR}{dt} \!=\! \frac{dR}{d\tau} \!=\!
   \left(\frac{1}{a}\frac{\partial R}{\partial t} + 
     \frac{1}{b}\frac{\partial R}{\partial r}
     \frac{b}{a}\frac{dr}{dt}\right) \!=\! 
     \left(U + \Gamma v \right) \ ,
\label{RWL2}
\end{equation}
 where
 \begin{equation}
 v\equiv \frac{b}{a}\frac{dr}{dt}\,,
\label{v}
\end{equation}
 is the 3-velocity of the object whose worldine is being considered, 
measured with respect to the local comoving frame of the fluid. 

For radial null rays, inserting $ds=d\Omega=0$ into the metric 
(\ref{eq_metric_MS}) gives $a^2 dt^2 = b^2 dr^2$, so that $v = \pm 1$, as 
expected, with the $+$ sign for an outgoing null ray and the $-$ sign for 
an ingoing one. Putting this into Eq.(\ref{RWL2}), and setting the right 
hand side to zero gives the condition for the trapping horizons as
 \begin{equation}
\left.\frac{1}{a}\frac{dR}{dt}\right\vert_{\pm} = \left(U \pm \Gamma 
\right) = 0 \ ,
\label{RNR}
\end{equation}
 with $\Gamma = -U$ for future horizons and $\Gamma = +U$ for past 
horizons, as above, which again then both lead to \mbox{$R=2M$.} 

Comparing Eq.(\ref{RNR}) with Eqs.(\ref{eq_theta+}) and 
(\ref{eq_theta-}) gives
 \begin{equation}
 \theta_{\pm} = \frac{2}{R} (U \pm \Gamma) = 
    \frac{2}{aR}\left.\frac{dR}{dt}\right\vert_{\pm} \ ,
 \label{eq_theta_dRdt}
\end{equation} 
 which connects the geometrical approach involving the expansion $\theta$ 
used in \cite{Binetruy:2014ela} with the fluid approach used in 
\cite{Miller:2014qaa}. Finally, note that the so-called apparent horizon 
of a black hole (a section of the future trapping horizon) is the outermost 
trapped surface for outgoing radial null rays while the trapping horizon 
for an expanding universe (which is a past trapping horizon) is foliated 
by the innermost past-trapped surfaces for ingoing radial null rays.

\section{Causal Nature and Horizon Velocity}
\label{section_causal}
\subsection{Causal Nature}
\label{section_alpha}
Within the geometrical approach, the causal nature of the horizons 
mentioned above can be determined using the ratio of the Lie derivatives 
of the expansion that changes sign ($\theta_{v}$) in the two radial null 
directions \cite{Dreyer:2002mx}. This ratio, often denoted by $\alpha$, 
is defined as follows
 \begin{equation} \alpha \equiv \frac{\mathcal{L}_{v} 
  \theta_{v}}{\mathcal{L}_{nv} \theta_{v}} \ , 
 \label{eq_alpha_defi}
\end{equation}
 where evaluation at the horizon is implicit. The parameter $\alpha$ is 
negative/positive for timelike/spacelike horizons respectively, and goes 
to zero or infinity for null horizons. Now relating these Lie derivatives 
to the operators $D_t$ and $D_r$ of the Misner-Sharp-Hernandez 
formalism, we get
 \[ \mathcal{L}_{+} \theta_v = \mathcal{L}_{k^a} \theta_v = k^a \partial_a 
 \theta_v 
 \!=\! \left( \frac{1}{a} \frac{\partial}{\partial t} 
 +\frac{1}{b} \frac{\partial}{\partial r} \right) 
 \theta_v \]

\[ \mathcal{L}_{-} \theta_v = \mathcal{L}_{l^a} \theta_v = l^a \partial_a 
 \theta_v
 \!=\! \left( \frac{1}{a} \frac{\partial}{\partial t} 
 -\frac{1}{b} \frac{\partial}{\partial r} \right) 
 \theta_v \]
  which gives 
 \begin{eqnarray}
 \mathcal{L}_{\pm} \theta_v = \left( D_t \pm D_r \right) \theta_v \ .
   \label{eq_Lie_plus/minus}
\end{eqnarray}
 Then for the black hole case, where $\theta_v=\theta_+\propto 
(U+\Gamma)$ we have
 \begin{eqnarray}
 \alpha = \frac{\mathcal{L}_{+} \theta_+}{\mathcal{L}_{-} \theta_+} =  
\frac{(D_t + D_r) \theta_+}{(D_t - D_r) \theta_+} =
 \left. \frac{(D_t U + D_t \Gamma) + (D_r U + D_r \Gamma)}{(D_t U + D_t 
 \Gamma) - (D_r U + D_r \Gamma)} \right\vert_H \ ,
 \label{alpha_BH}
\end{eqnarray}
 whereas in the case of an expanding universe, where 
$\theta_v=\theta_-\propto (U-\Gamma)$, we get
 \begin{equation}
 \alpha = \frac{\mathcal{L}_{-} \theta_-}{\mathcal{L}_{+} \theta_-} = 
 \frac{(D_t - D_r) \theta_-}{(D_t + D_r) \theta_-} =
\left. \frac{(D_t U - D_t \Gamma) - (D_r U - D_r \Gamma)}{(D_t U - D_t \Gamma) + 
 (D_r U - D_r \Gamma)} \right\vert_H \ .
 \label{alpha_CH}
\end{equation}
 Using the Misner-Sharp-Hernandez equations, these two expressions 
for $\alpha$ turn out to give the same result, valid for both black holes 
and cosmology
 \begin{equation}
 \fbox{ $\displaystyle{
 \left.\alpha=\frac{4\pi R^2(e+p)}{1-4\pi R^2(e-p)}\right\vert_H }$ } 
 \label{alpha}
\end{equation}
 where the $H$ indicates that the quantities are evaluated at the horizon 
location. (We note that this equation is equivalent to Eq.(2.7) of 
\cite{Booth:2007ix}, with the quantity $C$ used in that work being equal 
to $2\alpha$ here.)

In \ref{Appendix_MSH} we briefly introduce the Misner-Sharp-Hernandez 
equations, guiding the reader through the algebra to derive Eq.(\ref{alpha}).

\subsection{Horizon Velocity}
\label{section_vH}
 An alternative way of determining the nature and behaviour of the 
horizons is to calculate how they move with respect to the matter. This 
can be done by following the location where either of the defining 
conditions $\theta_v=0$ or $R=2M$ is satisfied. The two methods give 
the same result; for analogy with the derivation of $\alpha$, we choose 
also here to work with the expansion $\theta$. Maintaining 
$\theta_v=0$ along the worldline gives
\begin{equation}
  d\theta_v = \frac{\partial \theta_v}{\partial t}\,dt
    + \frac{\partial \theta_v}{\partial r}\,dr = 0 \,,
\label{HWL}
\end{equation}
 leading to
\begin{equation}
  D_t \theta_v + \frac{b}{a}\frac{dr}{dt} D_r \theta_v = 0
 \label{HWL2}
\end{equation}
 at the horizon, so that its 3-velocity with respect to the matter, using 
the general definition of $v$ given by Eq.(\ref{v}) evaluated here at the 
horizon location (cf. \cite{Booth:2007ix}), is then
\begin{equation}
  v_H \equiv - \frac{D_t\theta_v}{D_r \theta_v} \ .
\label{v_H}
\end{equation}
 This expression is somewhat analogous to Eq.(\ref{eq_alpha_defi}): 
in the hydrodynamical approach the proper time and proper space 
derivatives of the expansion $\theta_{v}$ are playing a similar role to 
that played in the geometrical approach by the Lie derivatives in the two 
null directions.
 
Differentiating through Eq.(\ref{eq_theta-theta+}) and setting $\theta_v 
= 0$, the expression for $v_H$ can be rewritten as
 \begin{equation}
 v_H = - \left.\frac{D_t \left(\Gamma^2-U^2\right)} 
  {D_r \left(\Gamma^2-U^2\right)}\right\vert_H \ , 
\label{v_H2}
\end{equation}
 or, by inserting into Eq.(\ref{v_H}) the sum and difference of the two 
expressions in Eq.(\ref{eq_Lie_plus/minus}), it can be written as
 \begin{equation} 
 v_H = - \left.\frac{\mathcal{L}_{+} \theta_v + \mathcal{L}_{-} \theta_v}
 {\mathcal{L}_{+} \theta_v - \mathcal{L}_{-} \theta_v}\right\vert_H \ ,
 \label{v_H_cov}
\end{equation}
 which leads to a direct relation between $\alpha$ and $v_H$ 
\begin{equation}
 v_H = \pm \frac{1+\alpha}{1-\alpha} \ ,
 \label{v_H/alpha}
\end{equation}
 with the plus for black hole formation and the minus for an 
expanding universe. Finally we can obtain an expression for $v_H$, 
analogous to Eq.(\ref{alpha}) for $\alpha$, either by substituting that 
expression for $\alpha$ into Eq.(\ref{v_H/alpha}) or by using 
Eq.(\ref{v_H2}) together with Eqs.(\ref{D_rM}) and (\ref{D_tM}) of 
\ref{Appendix_MSH}. This gives
 \begin{equation}
\fbox{ $\displaystyle{
v_H =  - \left. \frac{U}{\Gamma}\right\vert_H 
\left. \frac{1+8\pi R^2p}{1-8\pi R^2e}\right\vert_H  }$ } 
\label{v_H3}
\end{equation}
 from which $v_H$ can be calculated in terms of quantities evaluated at 
the horizon location; $-\left. \frac{U}{\Gamma} \right\vert_H = \pm 1$ 
for future (black hole) and past (expanding universe) horizons 
respectively. 

By definition, $v_H$ is positive for an outward-moving horizon and 
negative for an inward-moving one, with the ``motion'' here being 
measured with respect to the matter. In the following we refer to these 
respectively as \emph{outgoing} and \emph{ingoing} horizons, which does 
not always coincide with ``outer'' and ``inner'' horizons in the 
terminology of \cite{Hayward:1993wb}, as we will appreciate better in 
Section \ref{section_black_hole}. 

As mentioned earlier, because the horizon is a ``mathematical'' surface 
which does not correspond to any physical object, its velocity with 
respect to the fluid can be larger or smaller than the speed of light. 
More precisely, if $|v_H| > 1$ (with $c=1$) the horizon is spacelike, if 
$|v_H| < 1$ the horizon is timelike and if $|v_H| = 1$ the horizon is 
null.

Evaluating Eq.(\ref{RWL2}) at the horizon location, with $v = v_H$, gives 
the evolution of the circumferential radius of the horizon $R_H$, which 
we will be discussing in Section \ref{General_perspective}. We note that 
this expression is nothing other than the Lie derivative of the circumferential 
radius $R$ taken along the vector field
 \begin{equation} 
t^a = \frac{1}{a} \left( 1,  - \frac{\partial_t(U \pm \Gamma)} 
 {\partial_r(U \pm \Gamma)}, 0, 0 \right) \, ,
\label{tangent_vector}
\end{equation}
 which is tangent to the horizon (upper sign for black hole formation, 
lower sign for an expanding universe).

\section{Black Hole Horizons}
 \label{section_black_hole}

In this section, we implement the formalism and notation introduced in 
previous sections for studying the emergence and behaviour of trapping 
horizons during gravitational collapse to form black holes, focusing 
mainly on stellar collapse. While we will be using various 
simplifications, our eventual interest here is firmly focused on 
obtaining results having relevance for realistic astrophysical 
situations. We are considering here only ``standard matter'' for 
which $e$ is always positive and $p$ is non-negative, so that the null 
energy condition ($e+p\geq0$) always holds.

We are restricting attention here to spherically symmetric non-rotating 
models. While this is certainly an important restriction (and we note the 
very interesting related work by Schnetter \emph{et al.} 
\cite{Schnetter:2006yt} including rotation) the main features of interest 
for us are already present for non-rotating models. A number of the 
characteristic features seen in our present work confirm behaviour 
reported in \cite{Schnetter:2006yt}. The most important simplifications 
in our treatment are that we limit our matter model to representing an 
unmagnetized perfect fluid with a polytropic equation of state 
$p=K\rho^\gamma$, where $\rho$ is the rest-mass density and $K$ and 
$\gamma$ are constants (see \ref{Appendix_eqstate}). Note that, 
while simplified, this does include pressure effects in a meaningful way 
if suitable values are taken for $K$ and $\gamma$. The case $K = 0$ 
corresponds to idealised pressureless matter, often referred to as 
``dust''.

There have been important studies related to trapping horizons made in 
the context of pressureless matter. Some of this work has been 
extremely interesting (see, for example, \cite{Booth:2005ng} and 
\cite{Bengtsson:2013hla}) but pressure effects play an important role in 
real stellar collapse to form black holes and can seriously change the 
picture as far as our present considerations are concerned. Viewing dust 
as a pressureless particle gas, the particles would strictly need to have 
zero velocity dispersion, corresponding to zero temperature, in violation 
of the third law of thermodynamics. Using dust as an approximation for 
collapse calculations can be relevant under certain circumstances but 
requires care since any initial non-zero velocity dispersion will tend to 
be amplified during collapse. While pressure gradients (the first 
term on the right hand side of Eq.(\ref{Euler_eq}) in \ref{Appendix_MSH}) 
become ineffective in the advanced stages of collapse due to falling 
values of $\Gamma$, the other effects of pressure typically become large 
then and should not be neglected.

Historically, some basic features of the behaviour of trapping horizons 
during collapse can already be seen in the paper by May \& White 
\cite{May:1966zz}, published fifty years ago, where they presented 
results from computations made with their version of the 
Misner-Sharp-Hernandez equations in \ref{Appendix_MSH}, going beyond 
the treatment given in the earlier classic paper by Oppenheimer \& 
Snyder \cite{Oppenheimer:1939ue} by including non-zero pressure in the 
collapsing matter. We will start this section by briefly recalling the 
situation for the Oppenheimer-Snyder solution \cite{Oppenheimer:1939ue} 
for pressureless matter starting from constant density.

\subsection{Oppenheimer-Snyder collapse}
\label{section_dust}

\begin{figure}[t!]
\vspace{-1.2cm}
 \centering  
\includegraphics[width=7.5cm]{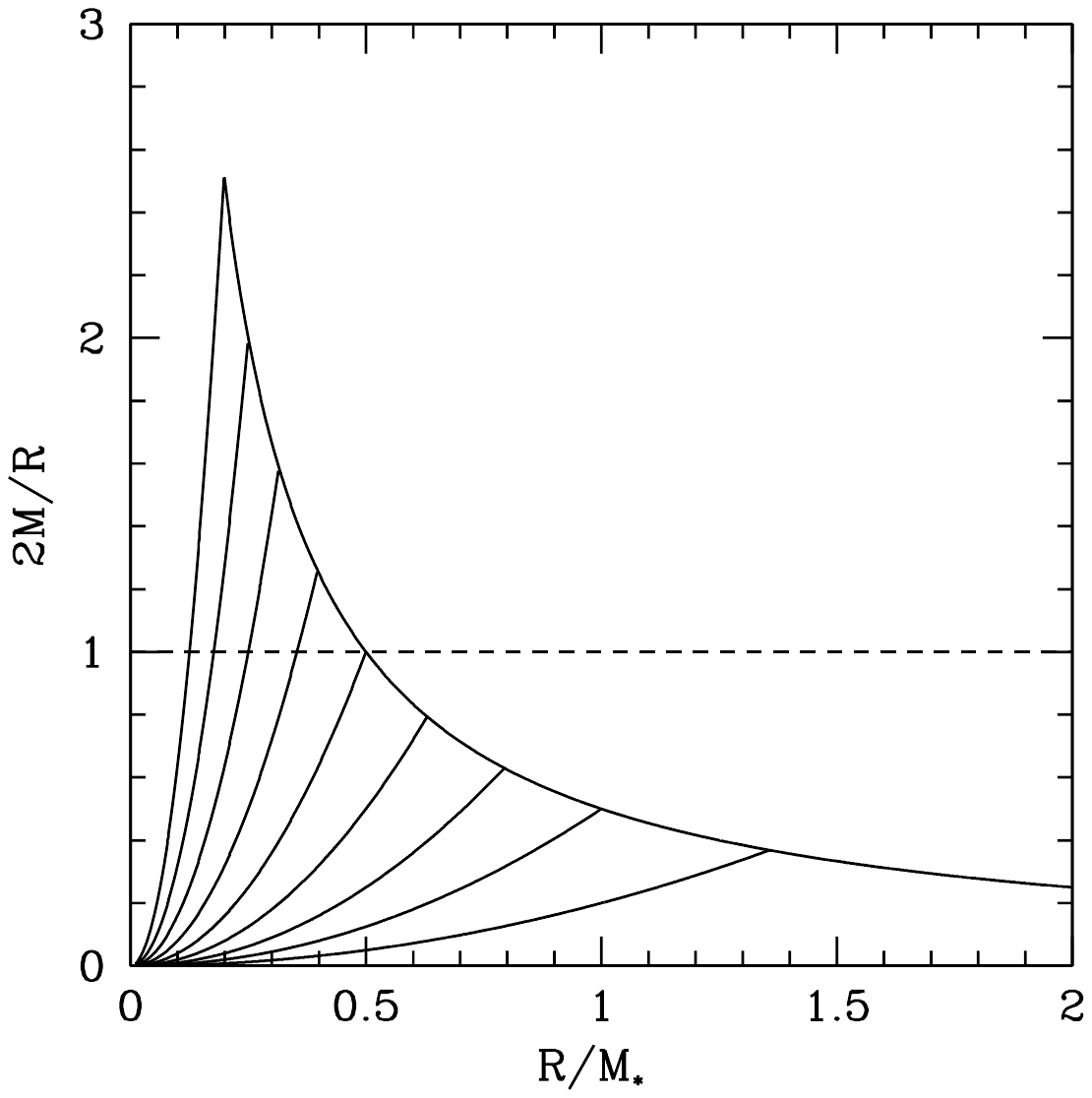} 
\includegraphics[width=6.5cm]{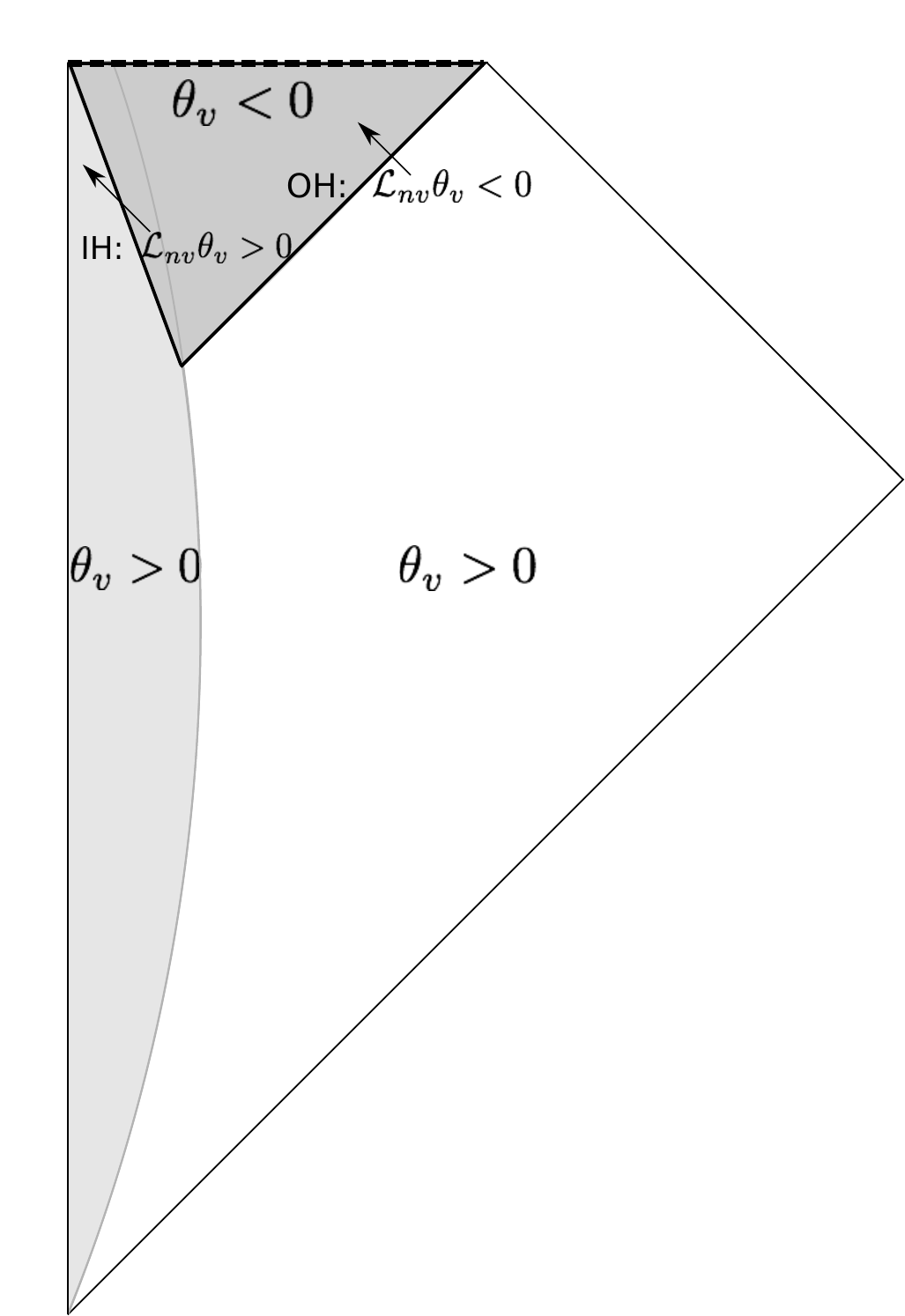}  
   \caption{Oppenheimer-Snyder collapse. (Left) $2M/R$ 
   plotted against $R/M_*$ (with $M_*$ being the total mass of the configuration) 
   at successive levels of the coordinate time $t$, with the time 
   increasing upwards. (Right) Penrose-Carter diagram for Oppenheimer-Snyder 
   collapse: the collapsing ball of pressureless matter is represented in light 
   grey, while the trapped region is shown in dark grey. The latter is 
   bounded by solid lines representing the horizons, with IH indicating 
   the \emph{inner} horizon and OH indicating the \emph{outer} one, 
   according to the classification of \cite{Hayward:1993wb} 
   ($\mathcal{L}_{nv}\theta_v>0$ and $\mathcal{L}_{nv}\theta_v<0$ 
   respectively, represented by the null arrows). Both OH and IH are here 
   \emph{future} horizons ($\theta_{nv}=\theta_-<0$). Note that in the 
   special case of Oppenheimer-Snyder collapse, OH is also 
   \emph{outgoing} ($v_H>0$) and IH is \emph{ingoing} ($v_H<0$).}
    \label{fig_OS}
\end{figure}

In the Oppenheimer-Snyder solution, one has zero pressure and 
uniform energy-density on each time level, with the singular condition 
being reached simultaneously by all of the collapsing matter in terms of 
the cosmic time used in Eq.(\ref{eq_metric_MS}) which, for pressureless 
matter, is equivalent to the proper time of comoving observers. (From 
Eq.(\ref{lapse_eq}) it follows that if $p=0$ the lapse is constant and 
can be normalised to $1$).

For this, we have
 \begin{equation}
M = \frac{4}{3}\pi R^3 e \ ,
\end{equation}
 so that 
 \begin{equation} 
\frac{2M}{R} = \frac{8}{3}\pi R^2 e \ , \label{2MR_OS} 
\end{equation} 
 from which it is clear that, since the energy density $e$ is a constant 
on each time slice, the condition $2M/R = 1$ is first reached at the 
surface of the configuration, where the energy density drops 
discontinuously from its uniform interior value to zero. Figure 
\ref{fig_OS} (left) shows the time evolution of the $2M/R$ curves, with 
the rising part of the curves corresponding to 
the matter configuration and the decreasing part to the vacuum outside. 
The horizons (occurring where $R = 2M$, which is shown with the 
dashed line) are here formed first at the surface of the configuration 
(at the gradient discontinuity), and the subsequent locations of the 
ingoing horizon correspond to successive intersections of the $2M/R$ 
curves with the dashed line. One should think of the ``outgoing'' horizon 
as being formed in the vacuum {\it outside} the configuration, where it 
is immediately null with
 \begin{equation} 
v_H = 1 \ , \quad\quad\quad \alpha = 0 \ , 
\end{equation}
and it then remains at a constant value of $R$; the ingoing horizon is 
formed just {\it inside} the matter with $8\pi R^2 e = 3$ so that
 \begin{equation}
v_H = - 1/2 \ ,  \quad \  \alpha = - 3 \ ,
\end{equation}
 values which continue to apply through all of its subsequent 
evolution\footnote{The particular value of $\alpha$ depends on the choice 
of normalisation, but the sign does not. In \cite{Booth:2005ng}, the 
normalisation gives $\alpha=-6$ whereas we have $\alpha=-3$.}. The fact 
of the horizons being born as null and timelike in this case comes from 
the very special nature of the Oppenheimer-Snyder solution, for which the 
birthplace of the two horizons coincides with the density discontinuity 
going from the non-vacuum interior to the vacuum exterior. One can here 
interpret the ingoing and outgoing horizons as two separate surfaces 
appearing at the same location, already having different values of 
$\alpha$ and $v_H$, whereas for all of the other cases studied here, 
the horizons emerge from the separation into two parts of a single 
marginally trapped surface which at the moment of formation is neither 
ingoing nor outgoing.
  
Figure \ref{fig_OS} (right) shows the Penrose-Carter diagram for this case, with 
IH and OH indicating the inner and outer horizons respectively. Note 
that, consistently with the behaviour in Figure \ref{fig_OS} (left), 
IH and OH do not join to form a single smooth hypersurface in this case, 
there being a discontinuity in the tangent vector where they join, 
corresponding to the density discontinuity at the stellar surface.

\subsection{Simulation I: the May \& White case} \label{section_MW} 
 In Figure 14 of the May \& White paper \cite{May:1966zz}, where $2M/R$ is 
plotted against their comoving rest-mass coordinate $\mu$ at a succession 
of times during the collapse, the peak local value of $2M/R$ can be seen 
rising to become equal to $1$ and then increasing further, leaving two 
locations where $2M/R = 1$ which then separate, one going outwards with 
respect to the matter and the other going inwards (i.e.\ outgoing and 
ingoing horizons in the sense explained in the Introduction). The feature 
of the horizons forming in pairs has subsequently been discussed in various 
contexts by many other authors (including 
\cite{Schnetter:2006yt,Jaramillo:2009zz,Jaramillo:2011zw}). We decided that 
it would be interesting to repeat their calculations more or less exactly, 
but adding on the concepts which we have been presenting here 
regarding trapping horizons. In this sense, the following discussion can be 
seen as an addendum to their paper.
    
In a subsequent article, \cite{May:1967} they gave a very full account of 
how their computations were made, and we followed their methodology with 
only minor modifications (apart from needing to implement excision, so as 
to be able to follow the outgoing horizon all the way to the stellar 
surface). They used a standard type of largely explicit Lagrangian 
finite-difference method, with the comoving grid consisting of a 
succession of concentric spherical shells. The comoving radial coordinate 
of the outer boundary of each shell was identified with the 
time-independent value of the rest-mass $\mu$ contained within the sphere 
having this as its surface (i.e.\ $r$ was set equal to $\mu$, giving 
$b=1/4\pi R^2 \rho$)\footnote{Note the distinction between the two 
measures of mass: the rest mass $\mu$ is a conserved quantity at any 
comoving location, whereas the corresponding Misner-Sharp-Hernandez
mass $M$ can vary.}. They used 200 of these grid-zones, each containing 
the same amount of rest-mass $\Delta\mu$.

The 1966 paper deals both with stellar collapses which ``bounce'' and 
with continued collapse. We deal with just the latter here, for which 
they presented numerical results for collapse from rest of a 
$\gamma=5/3$, $21 M_\odot$ polytropic sphere of matter having an 
initially uniform rest-mass density ($\rho_0 = 10^7 {\rm g/cm^3}$) and 
specific internal energy ($\epsilon_0/c^2 = 9.61 \times 10^{-7}$). (The 
$21 M_\odot$ refers to the rest-mass of the configuration but, at this 
initial density, the rest-mass is essentially equal to the gravitational 
mass.) We successfully reproduced their results with these settings, but 
for the discussion below we needed a higher resolution and so used 2000 
zones rather than 200. Also, we found that we could not get sufficiently 
clean results with their methodology if we started from a density as low 
as $10^7 {\rm g/cm^3}$ and so we started instead from $10^{11} {\rm 
g/cm^3}$ but with the same value of $K$. This gave results with only 
surprisingly minor differences from theirs.

\begin{figure} 
\vspace{-1.2cm}
 \centering
  \includegraphics[width=7.5cm]{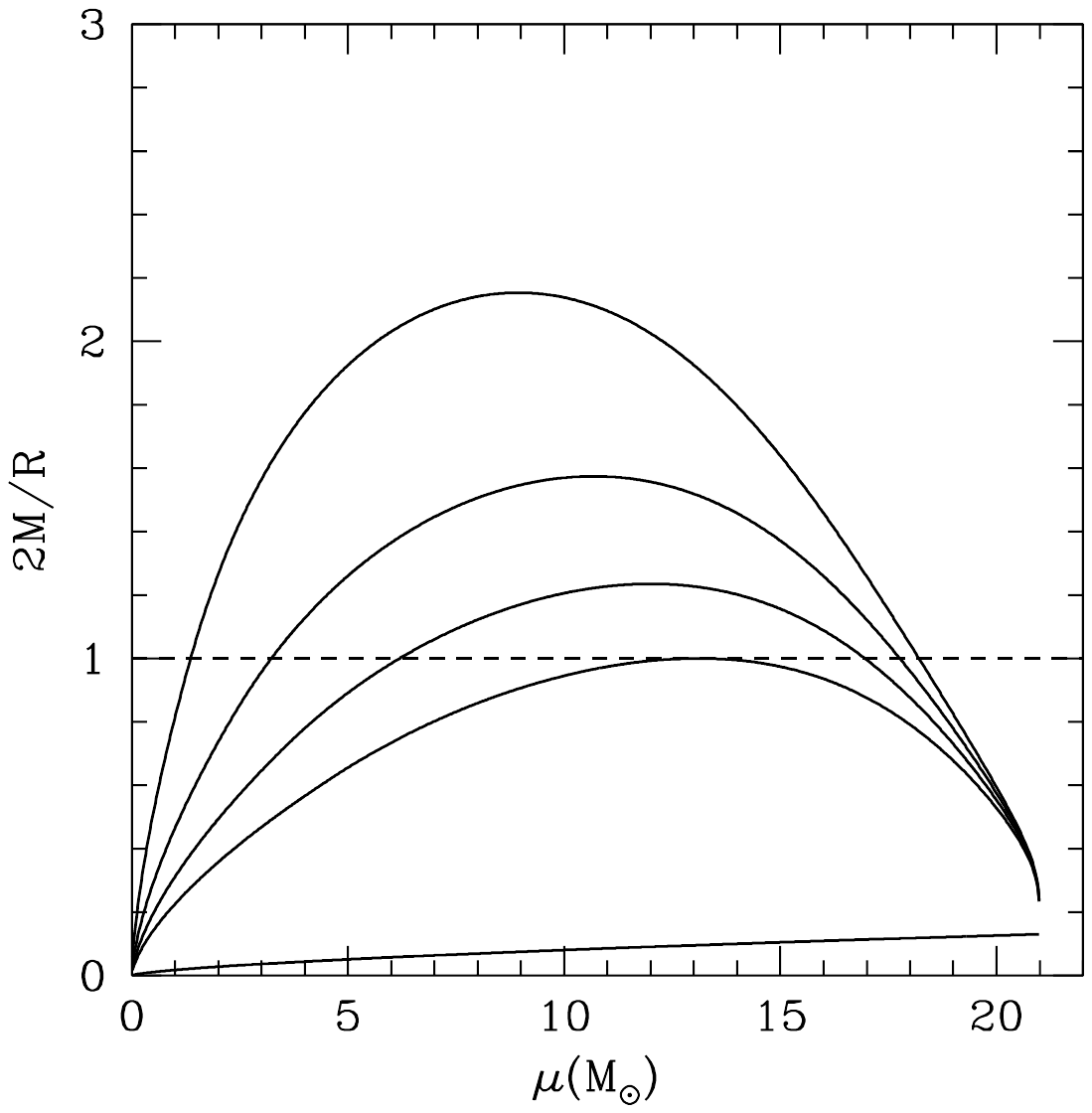}  \hspace {0.25cm}
  \includegraphics[width=7.5cm]{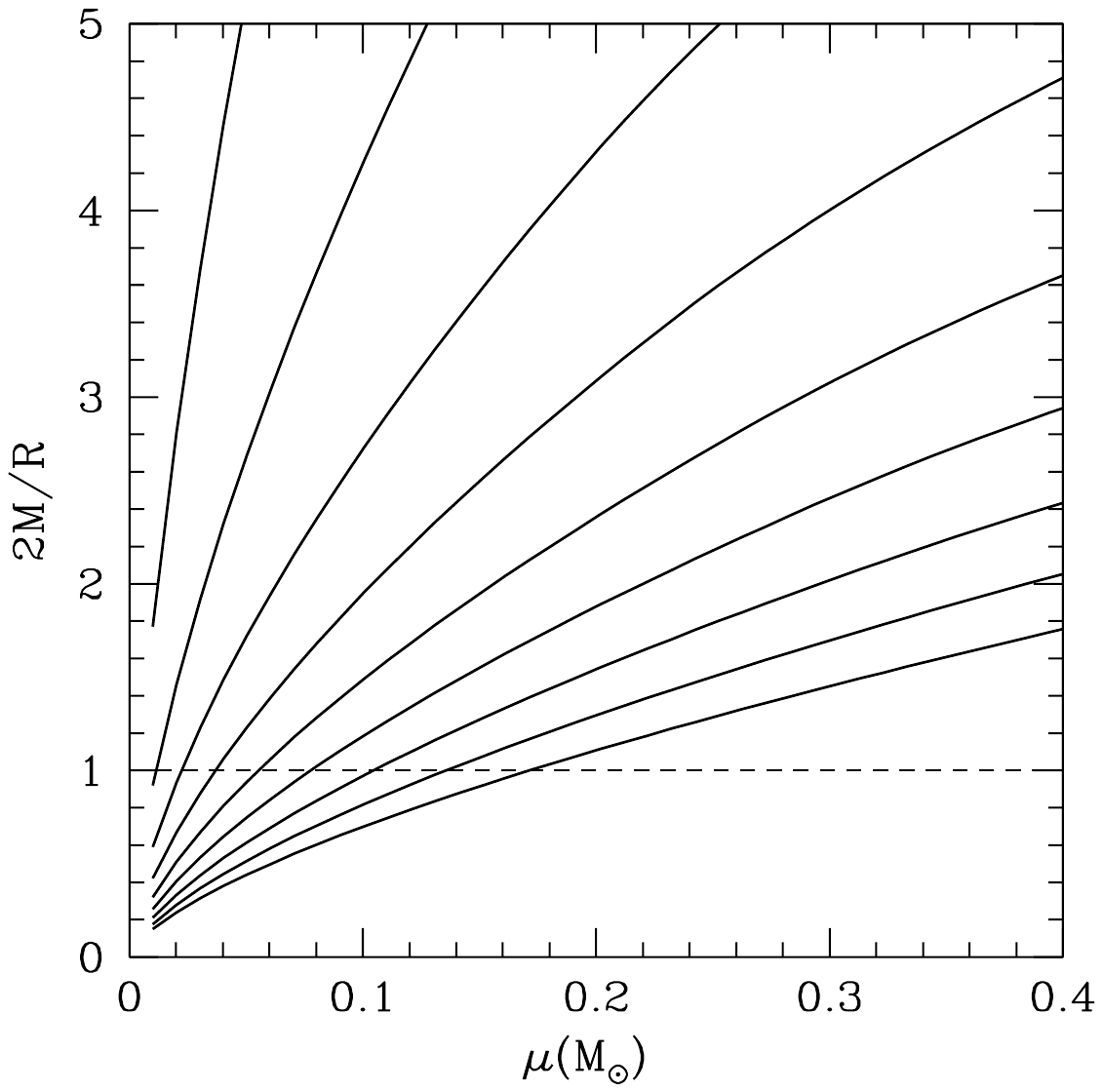} 
  \vspace{-2.0cm}
   \caption{$2M/R$ is plotted against $\mu$ at successive 
levels of the coordinate time $t$, with the time increasing upwards. This 
is for a collapse starting from a constant-density initial model with 
$\gamma = 5/3$. See text for further details.}
    \label{2m_R}
\end{figure}

The left hand frame of Figure \ref{2m_R} shows our plot of $2M/R$ 
versus $\mu$ corresponding to their Figure 14 but highlighting the points 
of interest to us here. The curves represent the solution at successive 
levels of the coordinate time $t$, with the time increasing upwards. (The 
total coordinate time for our run is two orders of magnitude smaller than 
May \& White's value, because of us starting from a higher density, but 
this is not relevant for our discussion.) Following the first (lowest) 
curve, representing the initial data, the next one up shows the moment when 
the curve first touches the condition $2M/R = 1$ (marked by the dashed 
horizontal line). At later times, there are then two points of intersection 
of the respective curves with the horizontal line, marking the locations of 
the ingoing and outgoing horizons which then move apart as time progresses 
further (cf. Figure 1 of \cite{Csizmadia:2009dm}). \mbox{Note that these 
are locations with respect to the {\it comoving coordinate} $\mu$.}
The right hand frame shows an expanded view of the eventual 
approach of the ingoing horizon to the inner edge of the grid. When it 
reaches the outer edge of the innermost zone (the last point shown) it can 
no longer be tracked with the present version of the code; we will 
return to discussion of the behaviour close to $\mu = 0$ at the end of this 
section.

\begin{figure} 
\vspace{0.5cm}
 \centering
  \includegraphics[width=9cm]{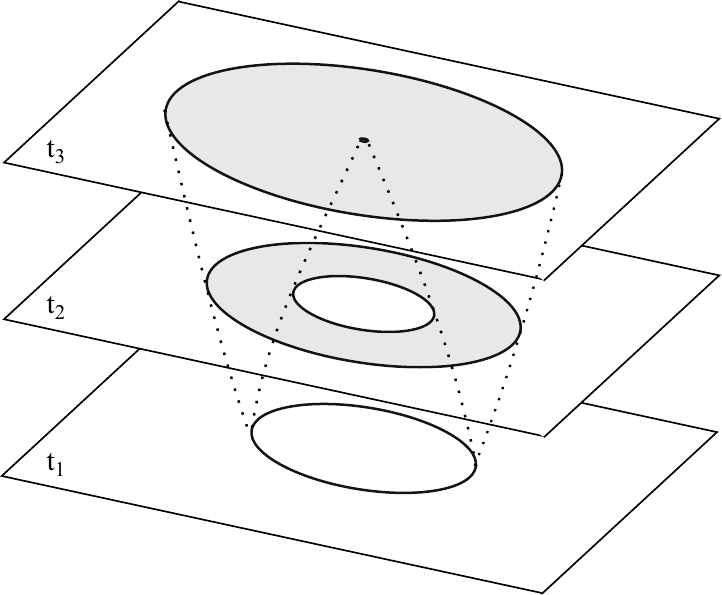}   
  \vspace{0.5cm}
   \caption{Schematic representation of the time-foliation of our spacetime. 
   The non-shaded parts of the drawing are the untrapped regions ($\theta_+>0$), 
   the shaded parts are the trapped regions ($\theta_+<0$), and the solid lines
   are the horizons ($\theta_+=0$). At first a single trapped surface appears, 
   which then separates into an outgoing horizon and an ingoing horizon. 
   Their evolution in terms of the radial coordinate $r=\mu$ is shown by the 
   dotted lines.}
    \label{schematic}
\end{figure}

These features are characteristic of collapses where the $R=2M$ condition 
is first reached within the bulk of the matter under conditions which are 
non-singular. Figure \ref{schematic} 
shows a schematic representation of the growth of the trapped region: 
initially a single horizon forms (at $t=t_1$) which then separates into 
ingoing and outgoing horizons which move apart ($t=t_2$) until eventually 
the ingoing horizon approaches the centre of symmetry ($t=t_3$) while 
the outgoing horizon approaches the outer edge of the 
matter and goes on to become an event horizon when it reaches the vacuum 
outside. The fact that the initial trapped surface should separate into 
ingoing and outgoing parts, rather than into two parts going in the same 
direction (with respect to the matter, as discussed above) follows from 
the fact that, during collapse of classical matter ($p\ge0$), the 
compactness $M/R$ calculated at constant values of $\mu$ is always 
increasing, as can be seen by calculating its time derivative:
 \begin{equation} 
D_t\left(\frac{M}{R}\right) = 
-\frac{U}{R}\left(\frac{M}{R}+4\pi R^2 p\right) \, ,
\end{equation}
 where Eq.(\ref{D_tM}) has been used. This is always positive when the 
matter is collapsing ($U<0$) and so the two locations where $R=2M$, 
arising after the separation, must then move in opposite directions 
(cf. Figure \ref{2m_R}).

\begin{figure}
\vspace{-1.2cm}
 \centering
 \includegraphics[width=7.5cm]{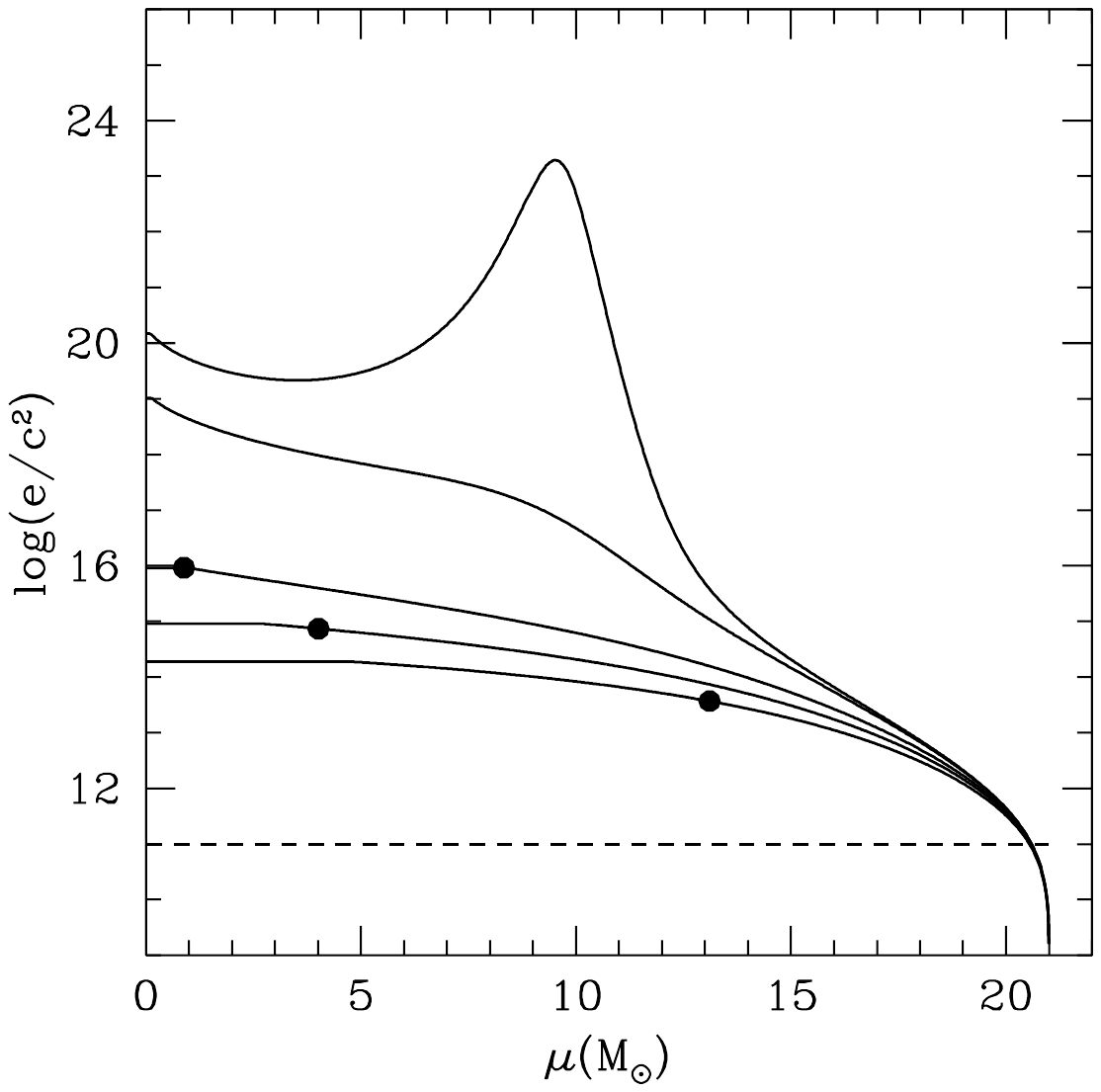}  \hspace {0.25cm}
 \includegraphics[width=7.5cm]{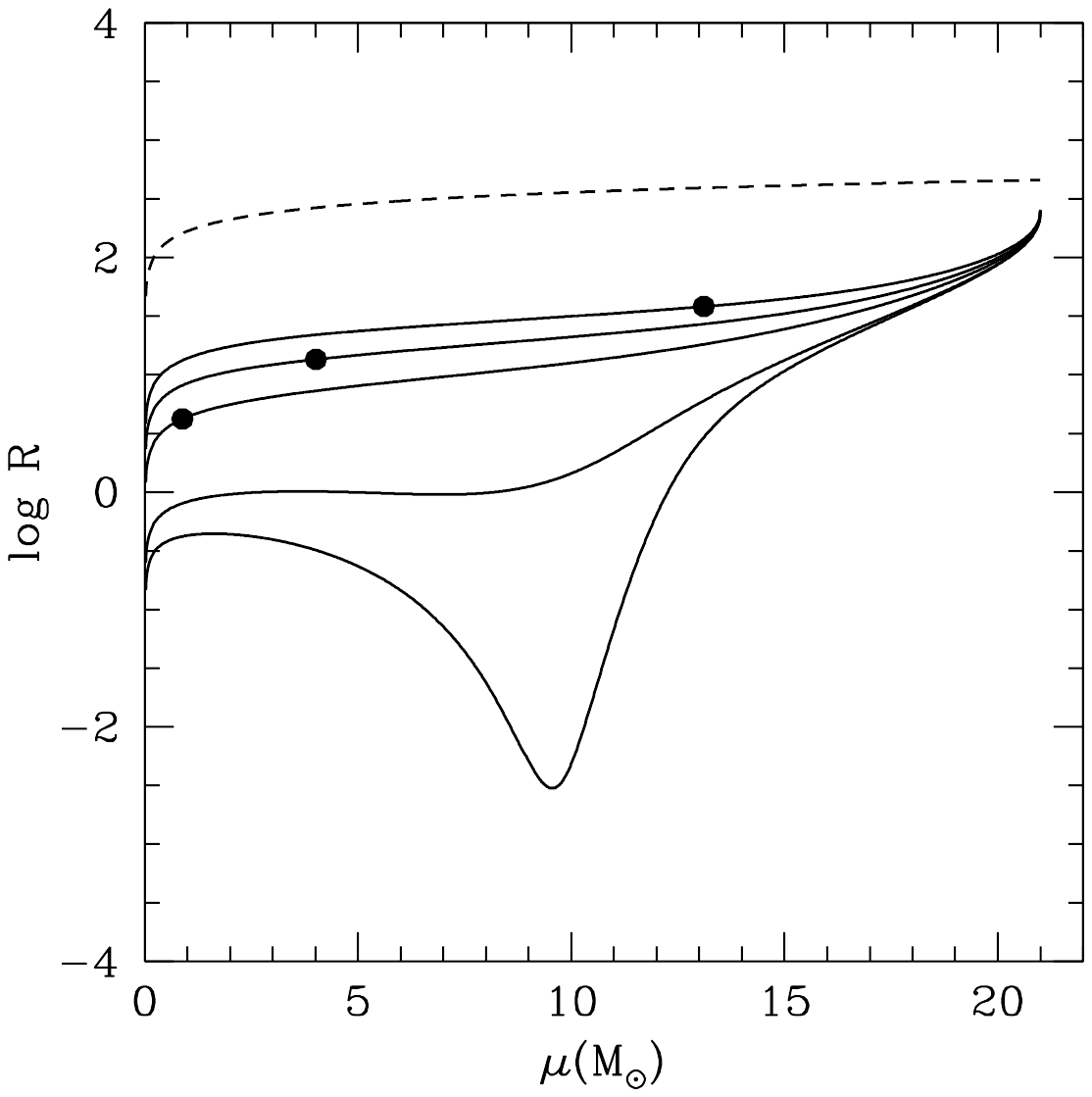}
 \vspace{-2.0cm}
   \caption{Plots of $e/c^2$ (in units of ${\rm g/cm^3}$) and $R$ 
(in units of km) against $\mu$ at successive levels of the coordinate 
time $t$, for the same case as in Figure \ref{2m_R}, with the time 
increasing upwards in the left-hand panel and downwards in the right-hand 
one. See text for further details.}
    \label{e&R_mu}
\end{figure}

In the left frame of Figure \ref{e&R_mu}, we have plotted the energy 
density $e$ as $e/c^2$ (in units of ${\rm g/cm^3}$) against $\mu$, for 
the same case as in Figure \ref{2m_R}. Here the horizontal dashed line 
marks the initial data, the next curve up is for the time at which the 
condition $2M/R = 1$ is first reached, and the successive curves are then 
for subsequent times (time increasing upwards). (May \& White plotted 
instead the {\it rest mass density} $\rho$, but it is the energy density 
which will concern us here.) The dots show locations of the {\it ingoing 
horizon} and will enter into our discussion below (the locations of the 
outgoing horizon are not shown here because they get confused among the 
closely-packed curves in the outer region). Initially there is a step 
discontinuity of the density (and pressure) at the surface of the 
configuration and this subsequently decays, giving a pressure gradient 
which works its way inwards. For a region where the pressure gradient 
remains zero, the density remains uniform (in this slicing), i.e.\ $e = 
e(t)$ with no dependence on $\mu$. Eventually (after the time of the 
third solid curve) the uniform region disappears completely and a density 
cusp appears at $\mu = 0$. Finally, an off-centred density maximum 
appears which goes on to form the curvature singularity, as was widely 
discussed in the 1960s (see \cite{Misner:1969br}, for example). The 
right-hand frame of Figure \ref{e&R_mu} shows a similar plot for $R$ (in 
units of km) versus $\mu$ (here time is increasing downwards). Note that 
at the curvature singularity, $R = 0$ (as expected) but $\mu \ne 0$ and, 
indeed, comprises a considerable fraction of the total rest mass. 
However, one needs to be cautious about this because the picture depends 
on the slicing used and while $\mu$ is defined in a clear physical way, 
the time coordinate $t$ is not, and depends on an arbitrary slicing 
choice. We are seeing here curves drawn at a succession of constant 
values of this coordinate $t$.

At this point, it is relevant to make some general comments (and we 
apologize to readers for whom this is well-known and obvious). General 
relativity is fully four-dimensional and reality resides in the full 
four-dimensional spacetime. It is a matter of convenience for us, in 
making calculations, to split the spacetime into 3 dimensions of space 
plus 1 of time. How this slicing is made is a matter of choice, and which 
events appear as simultaneous depends on this choice. Also, whether 
particular events in the four-dimensional spacetime appear or not with 
the slicing being used, depends on whether it reaches far enough to 
``see'' them. Because an event in the 4D spacetime is not ``seen'' by a 
given slicing does not mean that it does not exist. It is fundamental to 
the idea of a trapping horizon that it is a spacetime concept, defined 
quasi-locally, which does not depend on the asymptotic behaviour of a 
Cauchy slice\footnote{In this paper, we are dealing only with spherical 
symmetry and within that context trapping horizons are completely 
well-defined (in the sense explained in the introduction), although there 
may be ambiguities in other contexts.} 
\cite{Ashtekar:2004cn, Faraoni:2013aba}. Of the quantities which we are 
using here in our discussion, $R$ is invariantly defined (either as 
proper circumference divided by $2\pi$ or as an areal radius), $\mu$ is 
a physically-defined quantity, $b\,d\mu$ and $a\,dt$ are intervals of 
radial proper distance and proper time, as measured by observers 
comoving with the fluid, $v_H$ and $\alpha$ are defined directly in terms of 
$b\,d\mu$ and $a\,dt$, and $e$ and $p$ are measured in the fluid rest 
frame. It is true that working in the local comoving frame represents a 
choice, but comoving observers form a privileged class (see also \cite{Faraoni:2016xgy}).

\begin{figure} 
\vspace{-1.2cm}
 \centering
  \includegraphics[width=7.5cm]{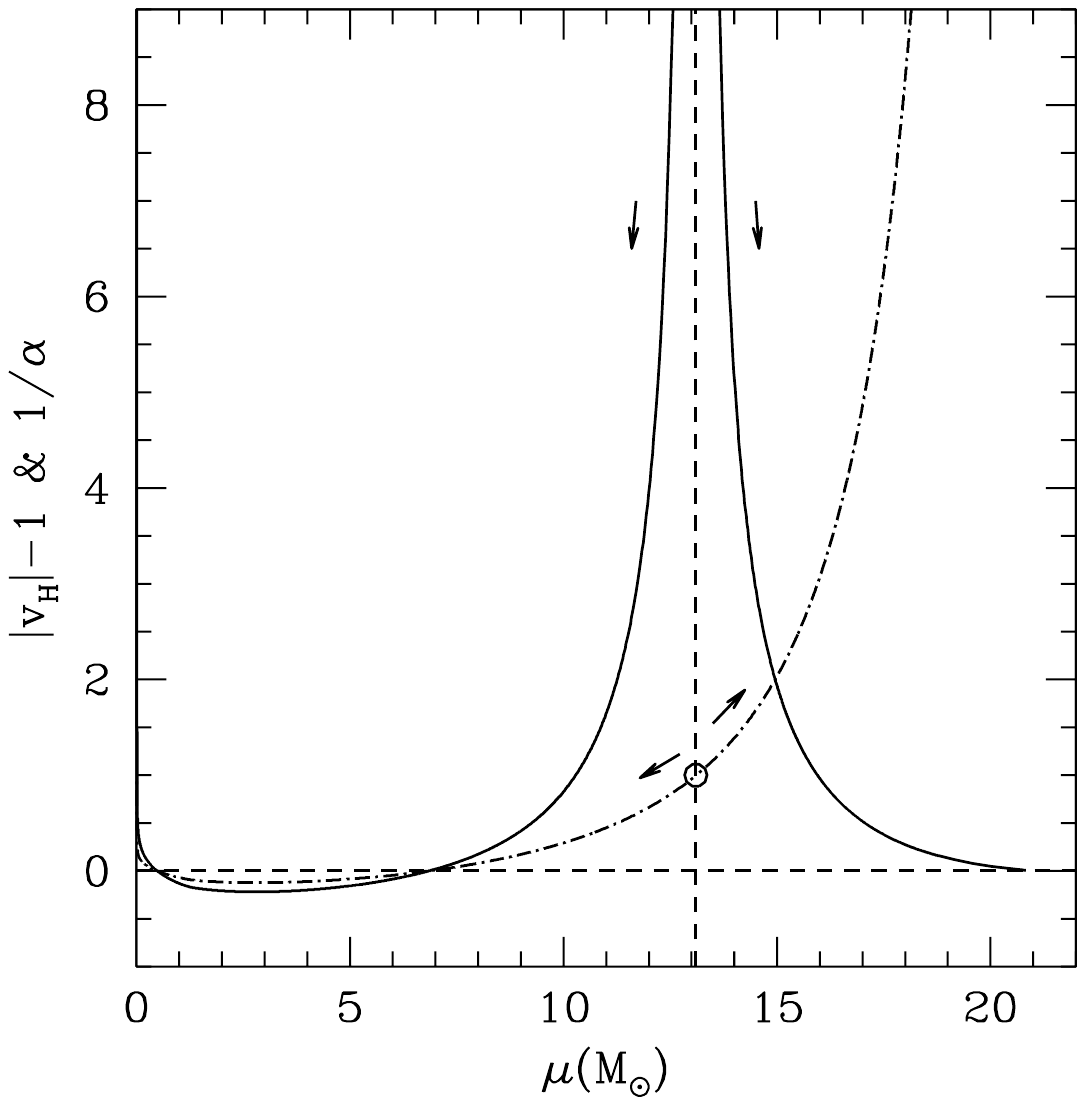} \hspace {0.25cm}
  \includegraphics[width=7.5cm]{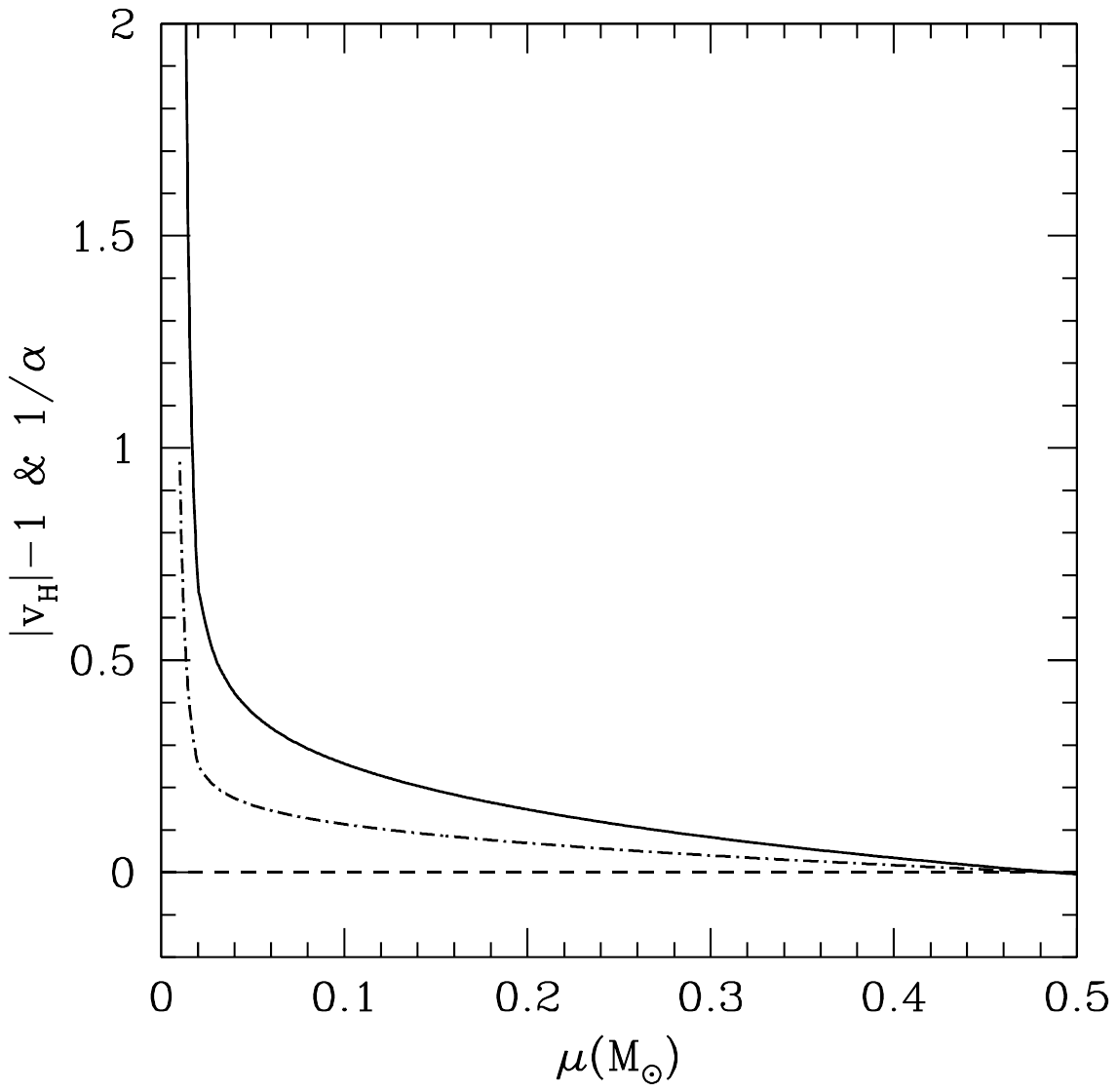} 
  \vspace{-2.0cm}
   \caption{Plots of $|v_H|-1$ (solid curves) and $1/\alpha$ (dot-dashed 
curves) as functions of $\mu$; this is for the standard $\gamma=5/3$ case 
for collapse from rest of a $21 M_\odot$ model with constant-density 
initial conditions. The right-hand frame shows an enlargement of the 
behaviour at small $\mu$. The arrows in the left-hand frame show the 
directions of time increasing along the horizon curves, the circle and vertical 
dashed line mark where the horizons form. }
    \label{vh_53_hom}
\end{figure}

Bearing in mind that Figures \ref{2m_R} and \ref{e&R_mu} need to be treated 
with some caution, we now proceed to discuss the behaviour of the trapping 
horizons using the quantities listed above. It is, of course, important to 
use a slicing for our numerical calculations in which the events which we 
want to study can be ``seen'', but the diagonal slicing used by May \& 
White does satisfy that. Figure \ref{vh_53_hom} shows the behaviour of the 
locally-measured horizon three-velocity $v_H$ and the parameter $\alpha$ 
for the outgoing and ingoing horizons from our standard run as described 
above (the right-hand frame shows an enlargement of the behaviour at small 
$\mu$). We have plotted the quantities $|v_H| - 1$ (solid curves) and 
$1/\alpha$ (dot-dashed curves); each of these is greater than zero if the 
horizon is spacelike, equal to zero if it is null, and less than zero if it 
is timelike. The behaviour seen can be summarized as follows. The horizons 
are born as a pair by the separation into two parts of the initial single 
marginally trapped surface formed when the $2M/R$ curve becomes tangent to 
$1$, as seen in Figure \ref{2m_R}. They begin to separate with infinite 
velocity, one going outwards with respect to the matter and the other going 
inwards. At the birth ($|v_H| \to \infty$), $\alpha = 1$; this can be 
understood from the fact that the first contact between the slicing and the 
3-horizon is tangential. It then follows from Eq.(\ref{v_H3}) that $8\pi 
R^2 e = 1$ at the point of birth, i.e.\ $e = 1/2A_H$ there, where 
$A_H$ is the area of the horizon - an intriguing result. Both horizons then 
slow down as they move, the outgoing horizon becoming asymptotically null 
as it approaches the surface of the configuration and becoming a static 
event horizon when it reaches the vacuum. (Note that $v_H$ is always 
measured with respect to the local infalling matter; the event horizon is 
outgoing and null with respect to local light cones but is static in that 
it remains at a constant value of $R$.) Meanwhile, $1/\alpha \to \infty$, 
so that $\alpha \to 0$ at the surface, as expected. As the ingoing horizon 
slows down, it becomes timelike for a while before becoming spacelike again 
and eventually going back to the conditions ($|v_H| \to \infty$, $\alpha = 
1$) which it had at birth. Comparing this with Figure \ref{e&R_mu}, note 
that the pair of horizons first forms at a value of $\mu$ outward of that 
for the singularity and that the uniform-density region is reached by the 
ingoing horizon while it is timelike. The rise of $|v_H|$ back towards 
infinity happens when the ingoing horizon encounters the rising part of the 
profile for $e$, leading towards the cusp at $\mu = 0$. All of the latter 
parts of this happen at values of $\mu$ smaller than that at which the 
singularity forms. As mentioned in connection with Figure \ref{2m_R}, 
we are not able to follow evolution of the ingoing horizon extremely close 
to $\mu = 0$ with our current methodology due to the finite resolution of 
the code: the last point shown for it in the figures is where it reaches 
the outer edge of our innermost zone, and $|v_H|$ has not yet diverged to 
infinity there ($\alpha$ is extremely close to $1$ but not coincident with 
it). It seems clear that the behaviour seen must be modified very close to 
$\mu = 0$ and we suspect that the ingoing horizon is stopped before 
reaching there, as we will discuss further in Section \ref{final_state}.

The behaviour seen in this simulation is rather different from that 
for the Oppenheimer-Snyder collapse described in the previous subsection, 
where the ingoing and outgoing horizons first appear already as separate 
entities. Another special feature of the Oppenheimer-Snyder solution is the 
one-to-one correspondence between outgoing/ingoing horizons and outer/inner 
horizons in the Hayward terminology, which does not arise for our 
simulations with non-zero pressure. Also, the combination of homogeneity 
and zero pressure preserves the density discontinuity at the surface which 
non-zero pressure instead removes due to the action of pressure gradients 
during the collapse. Note, however, that if one makes a dust calculation 
for models with non-uniform initial density profiles having no density 
discontinuity at the surface, such as the Lema{\^i}tre-Tolman-Bondi (LTB) 
models discussed in \cite{Booth:2005ng} (we have repeated those 
calculations), one again observes some cases where the horizons first 
emerge from a single trapped surface formed within the matter, which then 
separates into ingoing and outgoing horizons with a smooth join between 
them, reproducing the standard behaviour observed in Figure \ref{2m_R}. In 
those cases, the horizons again start with $\alpha=1$ and $|v_H|\to \infty$ 
but if the inner part of the initial density profile is nearly constant, 
one observes the Oppenheimer-Snyder values of $\alpha$ and $v_H$ being 
reached asymptotically as limiting values with no eventual return to the 
conditions at birth.

\begin{figure}[t!]
\vspace{-1.2cm}
 \centering
  \includegraphics[width=7.5cm]{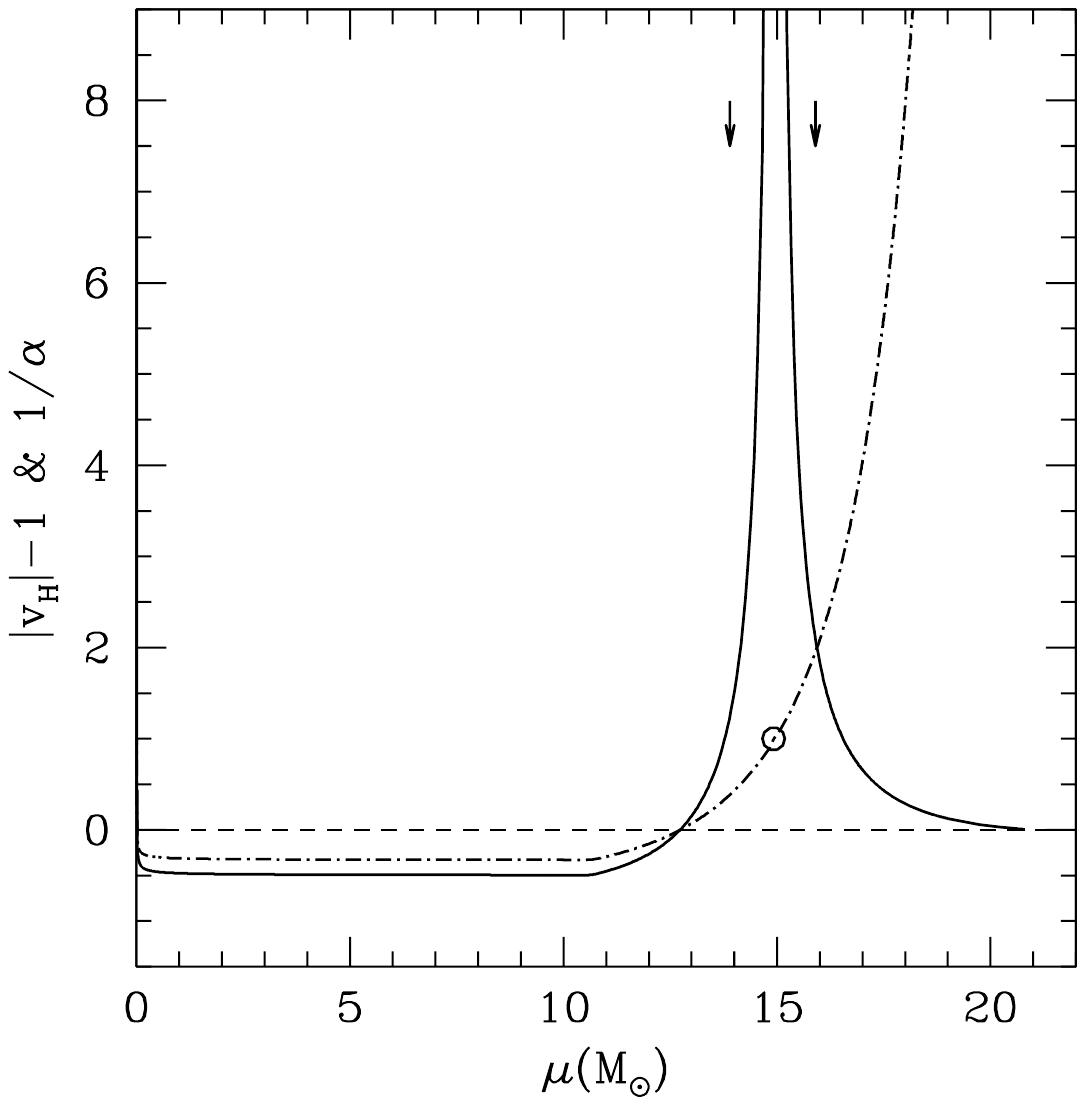} \hspace {0.25cm}    
  \includegraphics[width=7.5cm]{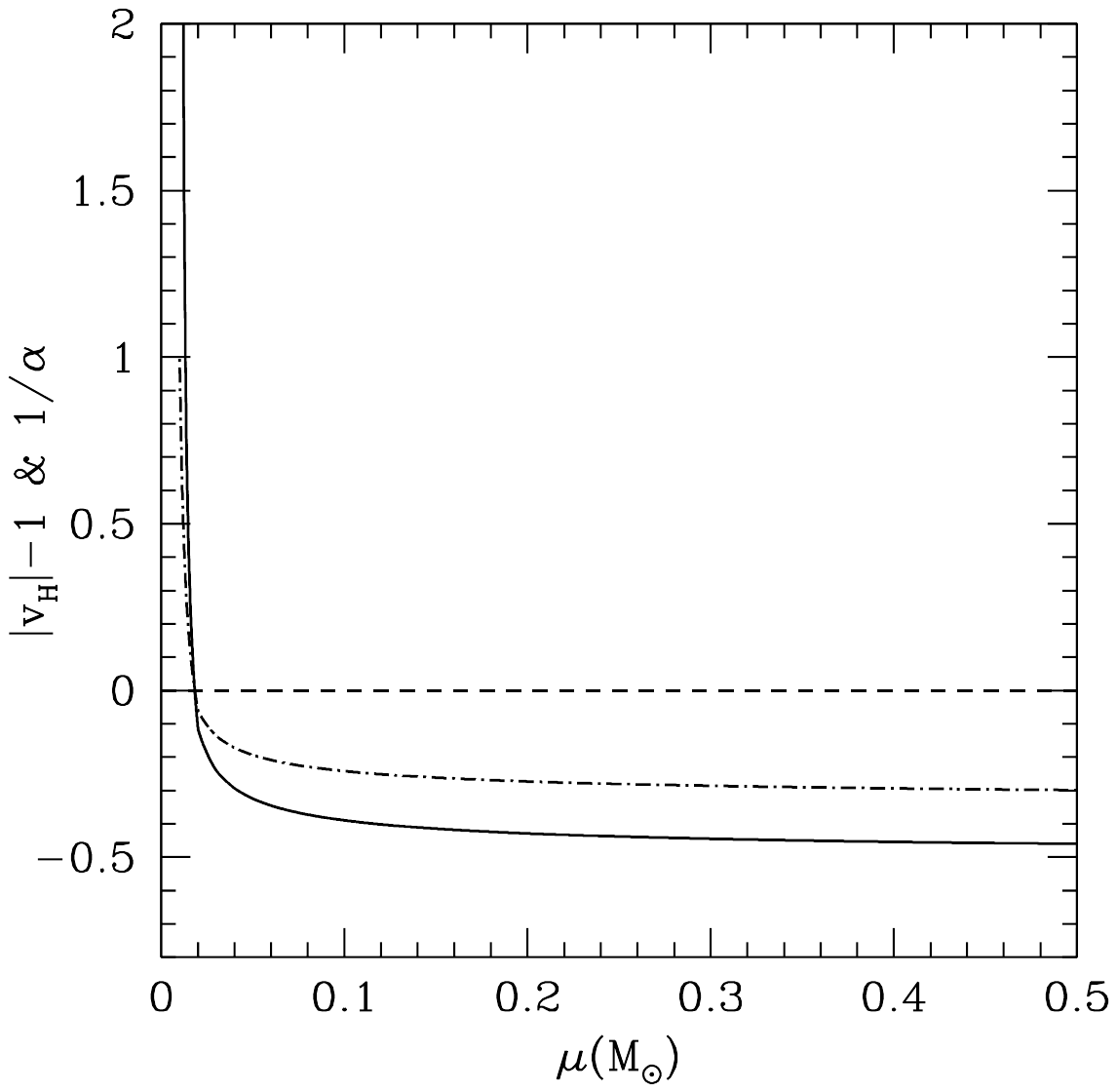}
  \vspace{-2.0cm}
   \caption{Equivalent plots to those in Figure \ref{vh_53_hom}, but this 
time with $\gamma=4/3$ rather than $5/3$. The solid curves again show 
$|v_H|-1$ while the dot-dashed curves show $1/\alpha$. The 
directions of time increasing along the horizon curves are the same as in 
Figure \ref{vh_53_hom} with the circle marking the location of horizon 
formation while the arrows show the directions of time increasing along the 
horizon curves.
}
    \label{vh_43_hom}
\end{figure}

\subsection{Simulations II \& III: further polytropic models}
\label{section_II_III}
 We wanted to investigate how the situation shown in Figure 
\ref{vh_53_hom} would change when the polytropic model for the 
configuration was altered. May \& White had chosen to use a model with a 
rest mass of $21 M_\odot$ with a $\gamma=5/3$ polytropic equation of 
state (representing a non-relativistic monatomic particle gas). For our 
first alternative model, we replaced $\gamma = 5/3$ by $\gamma = 4/3$, 
but kept everything else the same, including the initial value of 
$p/\rho$ (checking that this still ensured continuing collapse rather 
than a collapse and bounce). We chose $\gamma = 4/3$ (corresponding to a 
relativistic particle gas) because in the limit where the internal energy 
becomes very much larger than the rest-mass energy, this tends towards a 
$p=we$ equation of state with $w = 1/3$, as used for cosmological 
calculations in the radiative era of the early universe. The internal 
energy becomes progressively larger with respect to the rest-mass energy 
as the collapse proceeds but, in our calculation, it never quite reached 
the asymptotic regime. Our results are shown in Figure \ref{vh_43_hom}. 
The initial and final conditions for the horizons are similar to 
before; the most important difference seen is the long horizontal part of 
the curves for $|v_H| - 1$ and $1/\alpha$ when the ingoing horizon is 
timelike. The reason for this can be understood by looking again at 
Figure \ref{e&R_mu} (for $\gamma=5/3$). The horizontal regions of uniform 
energy density are longer in the $\gamma = 4/3$ case, so that the ingoing 
horizon is traversing spatially uniform matter for longer, meaning that 
there is a substantial part of the evolution which is similar to the 
Oppenheimer-Snyder solution \mbox{($p/e$ is quite small there).} In 
contrast, for the $\gamma = 5/3$ case the ingoing horizon only reached 
the uniform density region just before that disappeared.

\begin{figure}[t!]
\vspace{-1.2cm}
 \centering
  \includegraphics[width=7.5cm]{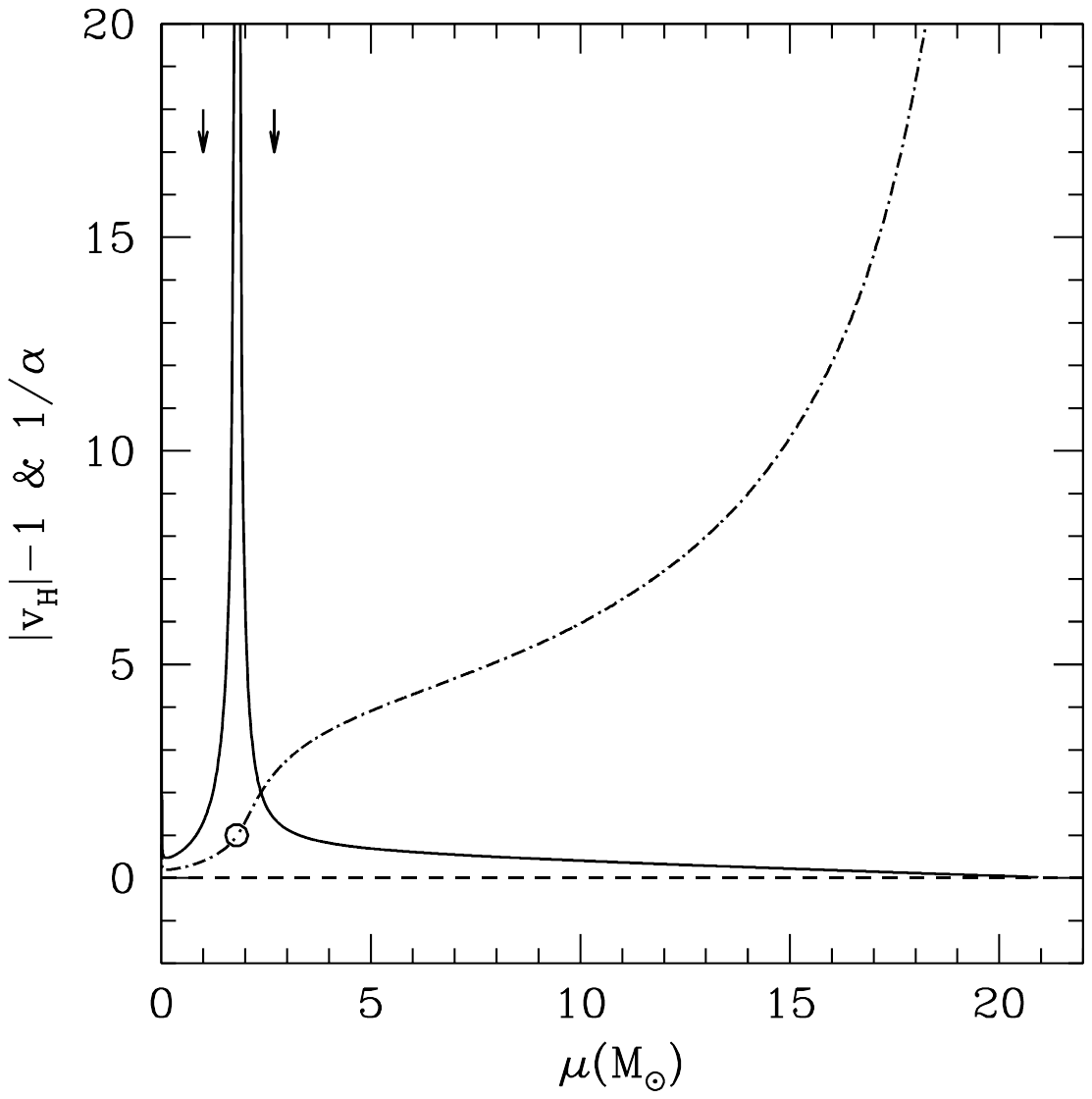}     \hspace {0.25cm}
  \includegraphics[width=7.5cm]{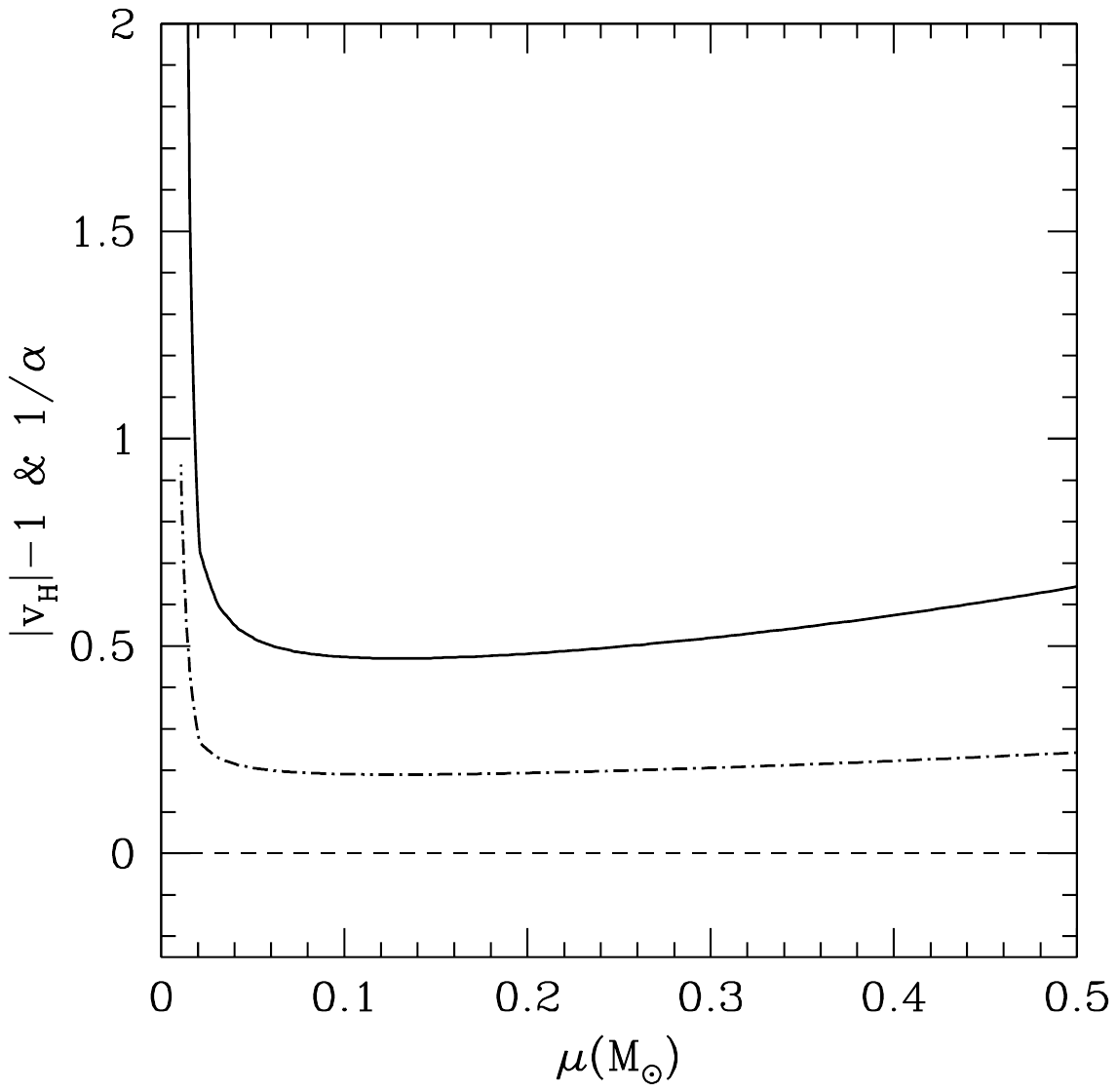}
\vspace{-2.0cm}
   \caption{Equivalent plots to those in Figure \ref{vh_53_hom}, using a 
$\gamma=5/3$ equation of state but starting from a $21 M_\odot$ model in 
hydrostatic equilibrium, which was then destabilized by decreasing $K$. 
The solid curves again show $|v_H|-1$ while the dot-dashed curves show 
$1/\alpha$. The directions of time increasing along the horizon 
curves are the same as before with the circle marking the location of horizon 
formation while the arrows show the directions of time increasing along the 
horizon curves.
}
    \label{vh_53_tov}
\end{figure}

The next trial returned to $\gamma = 5/3$ but abandoned constant density 
for the initial model, using instead a model in hydrostatic equilibrium, 
near to the radial instability limit, which was then made unstable to 
collapse by decreasing $K$. We again retained $21 M_\odot$ as the mass, 
for uniformity with the May \& White value. Results are shown in Figure 
\ref{vh_53_tov}.

Here, the pair of horizons are born at a much smaller value of $\mu$, 
because of the density being much higher in the central regions than 
further out, and there is never any uniform density region. In this case, 
the ingoing horizon is always spacelike (the reverse situation to the 
Oppenheimer-Snyder case). The other main features have been seen in all of the 
polytropic simulations which we have investigated here (horizons born as 
a pair, separating initially with infinite velocity and then slowing 
down; outgoing horizon being always spacelike but becoming asymptotically 
null as it reaches the stellar surface; ingoing horizon always being 
spacelike at the end and tending to infinite velocity close to $\mu=0$).

\begin{figure}
\vspace{-1.5cm}
 \centering
  \includegraphics[width=7.5cm]{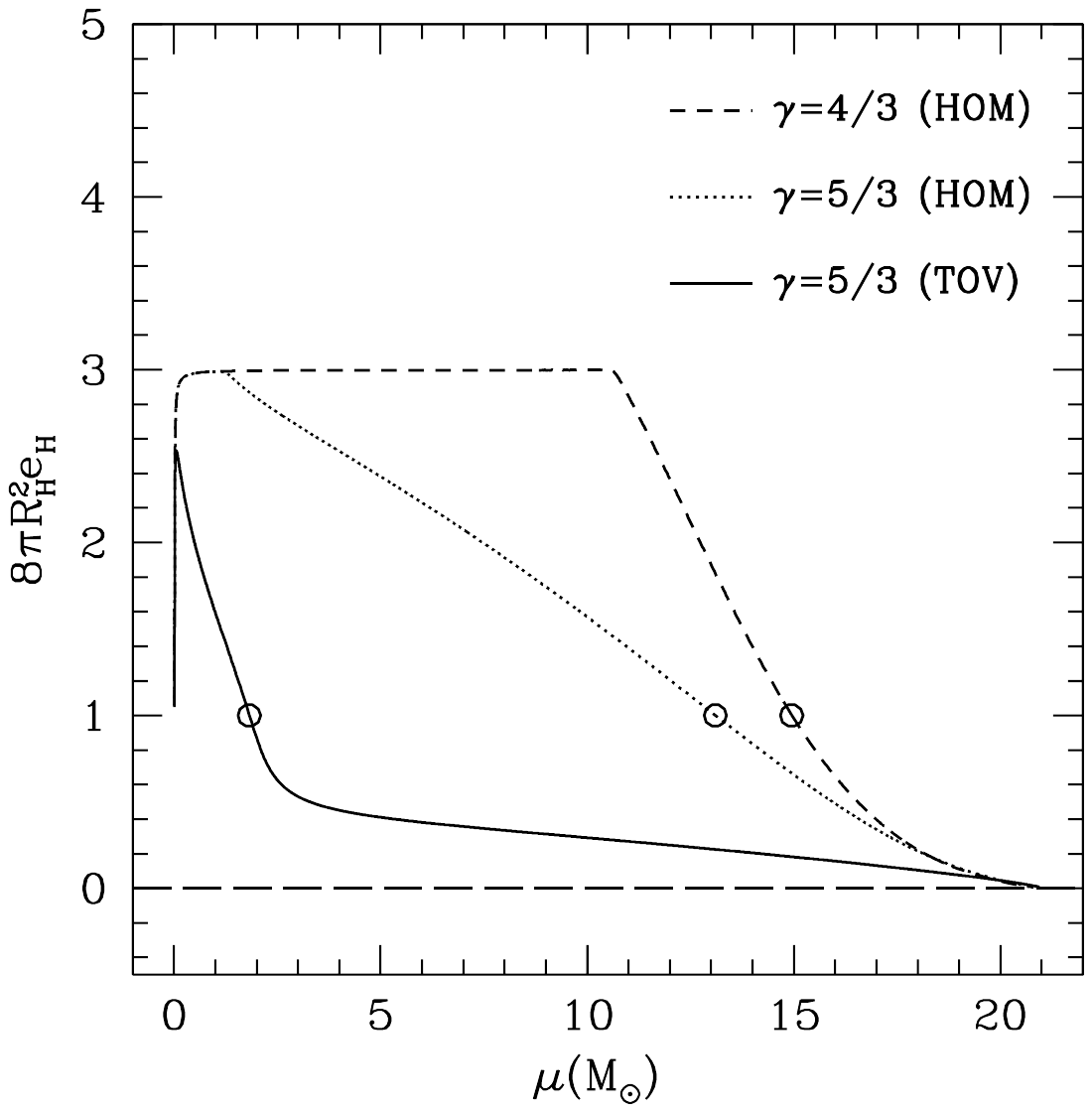}    
  \vspace{-2.5cm}
   \caption{The quantity $8\pi R_H^2 e_H$ is plotted against $\mu$ for 
each of the three cases studied. HOM indicates cases where the initial 
model was homogeneous (i.e.\ had constant density); TOV indicates that 
the initial model was obtained by solving the TOV equations of 
hydrostatic equilibrium and then making it unstable to collapse by 
reducing $K$. The circles indicate the join between ingoing and outgoing 
horizons.}
 \label{R2e_mu}
\end{figure}

Figure \ref{R2e_mu} shows $8\pi R^2 e$ (evaluated at the horizon 
location) plotted as a function of $\mu$ for each of the polytropic
simulations. This is the key quantity appearing in the denominator of 
Eq.(\ref{v_H3}). The circles mark the join between the results for the 
ingoing and outgoing horizons, at the point where they form (with 
$8\pi R^2 e = 1$ as mentioned earlier). Note the long horizontal section 
at the Oppenheimer-Snyder value of $8\pi R^2 e = 3$ for the $\gamma = 4/3$ 
case starting from constant density, while this value is only barely 
reached for the corresponding $\gamma = 5/3$ case. For the run starting 
from hydrostatic equilibrium, $8\pi R^2 e$ at the horizon peaks at a value 
less than $3$. In the two runs starting with a homogeneous profile, 
the horizon is forming closer to the surface of the star and the ingoing 
horizon is able to become timelike, eventually reaching the homogeneous 
condition $8\pi R^2 e = 3$, whereas in the run starting from hydrostatic 
equilibrium the horizon forms much closer to the centre and the ingoing 
horizon remains always spacelike. In all of the cases, there is eventually 
a rapid fall towards $1$ as the horizon enters the region of increasing 
density leading up to the cusp.

\subsection{LTB models and the final state of the ingoing horizon}
\label{final_state}

\begin{figure}[t!]
\vspace{-1.2cm}
 \centering
\includegraphics[width=7.5cm]{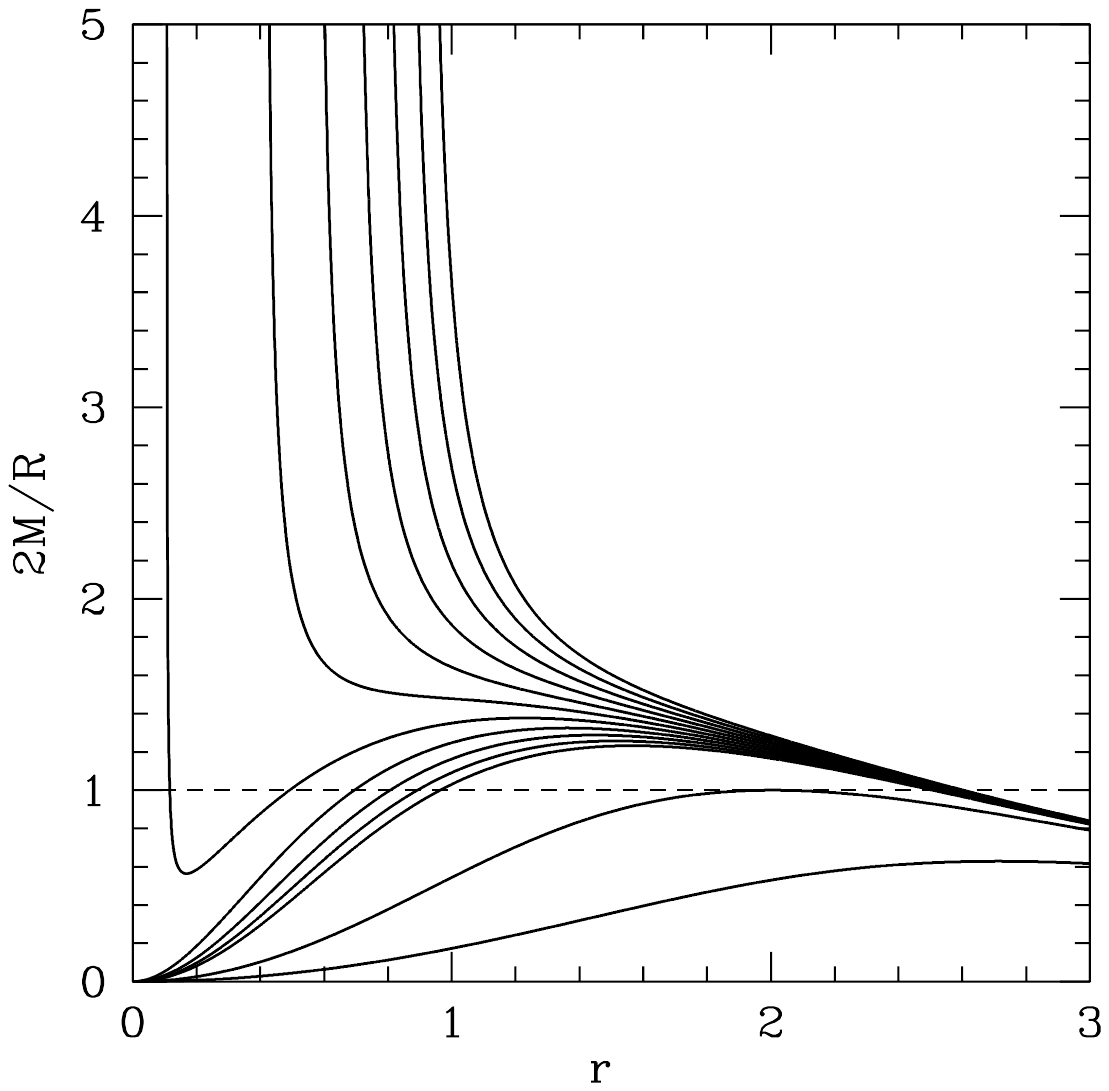}    \hspace {0.25cm}
  \includegraphics[width=7.5cm]{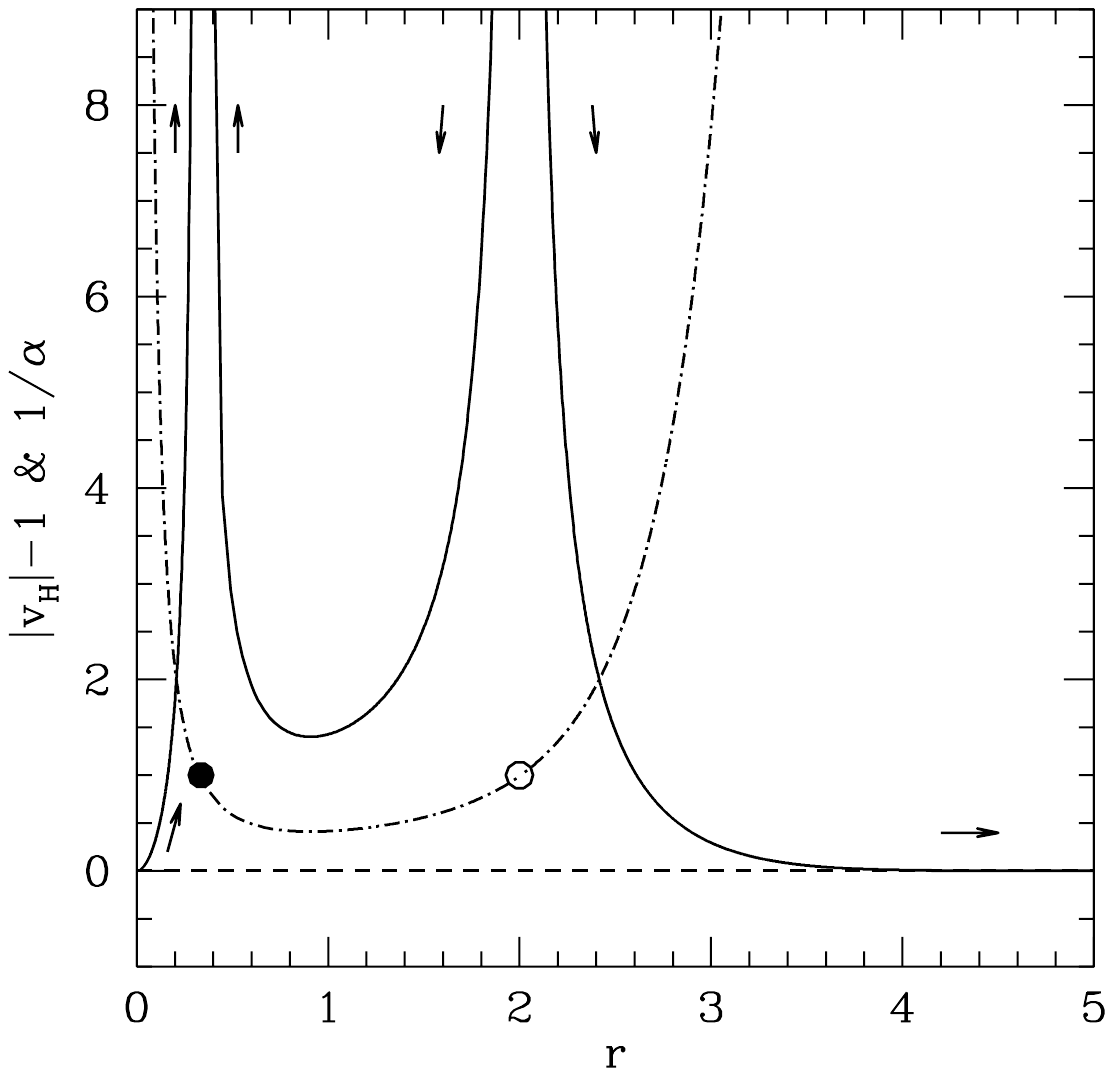}
  \vspace{-2.0cm}
 \caption{The left-hand frame shows $2M/R$ plotted against $r$, at 
  successive values of the comoving proper time, for the LTB model of 
  \cite{Booth:2005ng} with $\sigma=2$. The right-hand frame shows the 
  corresponding behaviour for $\alpha$ and $v_H$ with the solid 
  curves representing $|v_H|-1$ and the dot-dashed curves 
  representing $1/\alpha$. The directions of time increasing along the 
  horizon curves are the same as in Figure \ref{vh_53_tov} with the 
  addition of a secondary outgoing horizon forming at $r=0$ which 
  moves outward along the $\alpha$ and $v_H$ curves until it meets 
  and annihilates with the ingoing horizon at the inner location 
  where $v_H$ diverges. }
    \label{vh_sigma=2}
\end{figure}

If $|v_H| \to \infty$ (with $\alpha=1$ and $e_H=1/2A_H$) is genuinely 
the ``final'' condition for the ingoing horizon, as seems to be indicated 
by each of our three simulations, the fact that $e_H$ is not diverging 
there (see Figure \ref{e&R_mu}) indicates that $A_H$ should also be finite 
at the end and, if so, the ingoing horizon must be stopped before reaching 
$\mu = 0$ (unless $A_H$ could somehow be non-zero there). Our simulations, 
made following the methodology of May \& White \cite{May:1966zz}, are 
limited in seeing very small-scale phenomena by the use of a finite-spaced 
grid, but we thought it useful to look for hints about the possible 
eventual behaviour of the ingoing horizon from the quasi-analytic LTB dust 
models. As mentioned previously, the very interesting LTB calculations 
with smooth non-uniform density profiles presented in \cite{Booth:2005ng} 
include cases where the initial behaviour is rather similar to that seen 
in our simulations, and there is one case where the ingoing horizon 
is indeed  beingstopped before reaching the centre (although this is not 
commented on in their paper). In Figure \ref{vh_sigma=2}, we show results 
from our re-calculation of this case in terms of the quantities which we 
are using in this paper, apart from following the previous authors' use as 
a comoving coordinate of the circumferential (areal) radius of each shell 
at the initial time, which they denote by $r$. The model concerned is the 
$\sigma = 2$ model from Section 3.3.2 of their paper, and we refer the 
reader to \cite{Booth:2005ng} for further details of this and other 
associated models. Note that the use of $r$ as a radial coordinate 
effectively expands the innermost regions and contracts the outer ones 
with respect to what one would see using our mass coordinate $\mu$.

In the left-hand frame, we have plotted the behaviour of $2M/R$ at 
successive levels of the comoving proper time (with time increasing 
upwards). The lowest curve is for the initial time, the next one up is at 
the time when $2M/R$ first becomes equal to $1$ and then the following 
curves are at equal time intervals around the time when the ingoing 
horizon is stopped. The ingoing and outgoing horizons (located at the 
points where $2M/R=1$) initially form together inside the matter and then 
separate, as seen in the simulations, but the ingoing horizon eventually 
meets a secondary outgoing one emerging from $r=0$, where a singularity 
has formed, and annihilates with it, leaving only the outer outgoing 
horizon. The right-hand frame shows the corresponding behaviour of $|v_H|$ 
and $\alpha$, with the arrows marking the direction of motion along the 
$|v_H|$ curves. The two main horizons form as a pair at $r \simeq 2$ where 
$|v_H|$ diverges and $\alpha=1$ (marked with an open circle). They then 
separate as usual, both being spacelike. The velocity $|v_H|$ of the ingoing 
horizon first decreases but reaches a minimum velocity (still spacelike) 
and then increases again towards the left-hand divergence at $r\simeq0.33$. 
There it approaches the inner outgoing horizon and annihilates with it at 
the point where $\alpha$ is again equal to $1$ (filled circle). The processes 
of formation and annihilation appear as identical here, only with opposite 
directions of the arrow of time.

It can be seen that this reproduces some of the key points of 
the phenomenology seen in the simulations: the horizons forming together 
with $|v_H| \to \infty$ and $\alpha = 1$ and then separating, followed by 
the ingoing horizon ending again with the same conditions. However, 
other aspects (including the nature of the singularity formation) are 
significantly different because of pressure effects and associated changes 
in the time slicing. It remains to be seen the extent to which the process 
described above for the stopping of the ingoing horizon, corresponds to 
what occurs in physical cases where there is non-zero pressure.

\section{Causal Nature: a general perspective}
\label{General_perspective}
\begin{figure}
 \centering
  \includegraphics[width=13.5cm]{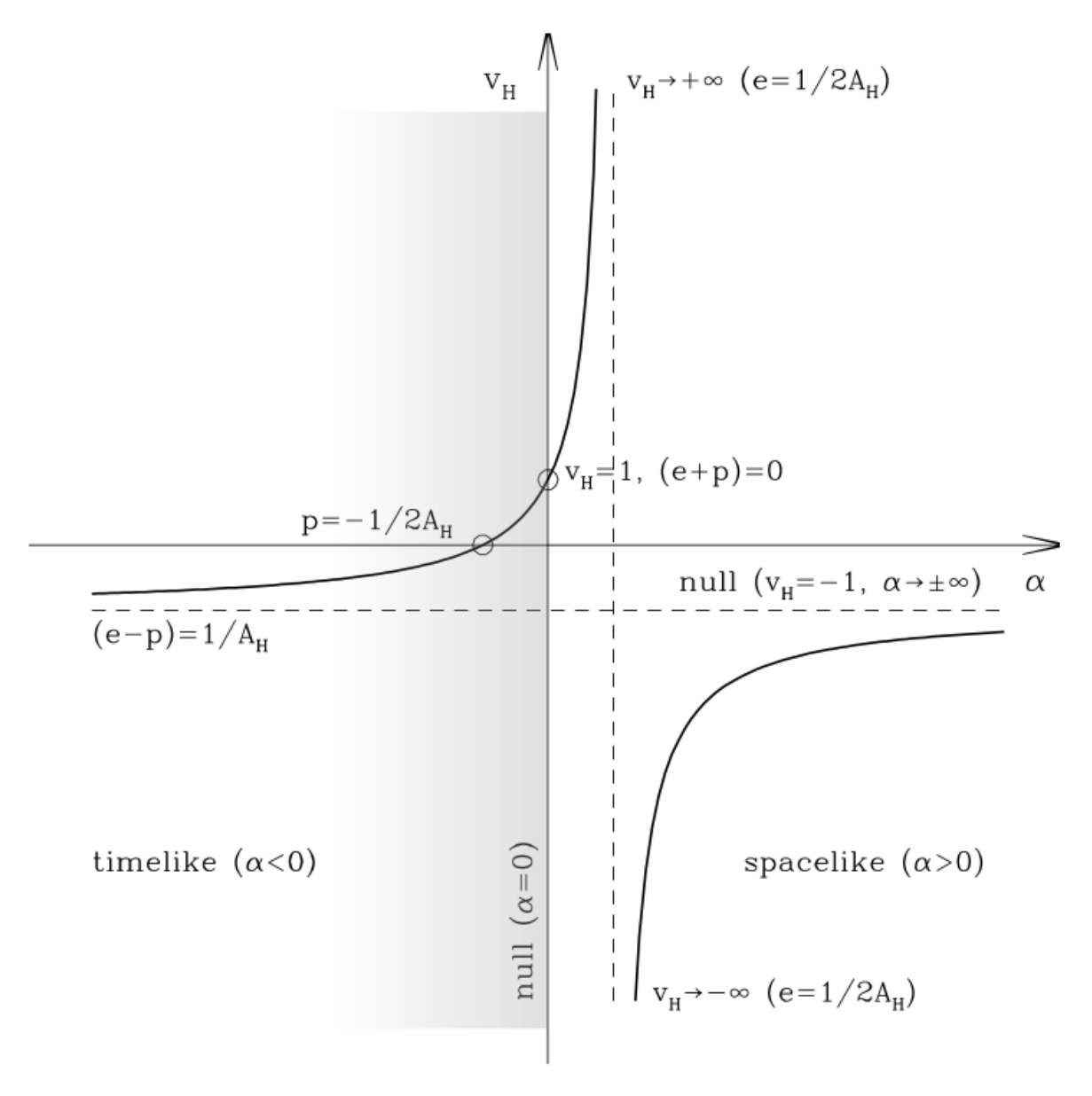}    
   \caption{Plot of $v_H$ versus $\alpha$ for the black hole horizons. The 
   significance of the points indicated in the figure is described in the 
   text.}
    \label{fig_BH(alpha-v_H)}
\end{figure}

 Here we give a more general overview of the types of behaviour which we 
have observed in the numerical simulation results, summarising the 
different phenomenologies within a coherent picture. We stress that 
everything in this section refers to strictly spherically symmetric 
horizons within a spherically symmetric spacetime. First we recall that 
for black hole trapping horizons, it is the outward expansion 
$\theta_{+}$ which vanishes, these being \emph{future} horizons in the 
Hayward terminology, with $\Gamma=-U$ at the horizon location. 
As we show in the following, it is useful to exhibit the simulation 
results on a plot of $v_H$ against $\alpha$, since this allows 
identification of all of the conceivable horizon behaviours within 
the scenario being considered here, together with indicating 
the corresponding conditions for energy density and pressure. 
Using Eq.(\ref{v_H/alpha}) written for the black hole case, we have
 \begin{equation}
  v_H = \frac{1+\alpha}{1-\alpha} \,,
\label{eq_alpha_vH_BH}
\end{equation}
which is represented by a rectangular hyperbola, as shown in Figure 
\ref{fig_BH(alpha-v_H)}. This expression is very general, based only on 
the definitions of these quantities in spherical symmetry (see Section 
\ref{section_causal}), and does not depend on the particular form of the 
equation of state or of the stress energy tensor. It links 
together the geometrical approach to the causal nature, represented by 
$\alpha$, and the hydrodynamical one, represented by $v_H$, corresponding 
to the two sides of the Einstein equation. Although $\alpha$ and $v_H$ are 
not independent quantities, we suggest that discussing both of them together 
is useful for clarifying the physical meaning of some key features, creating 
a common ground of understanding between the geometrical and 
hydrodynamical approaches.

For collapse of a perfect fluid $\alpha$ and $v_H$ are given by
 \begin{equation}
 \alpha=\frac{4\pi R_H^2(e+p)}{1-4\pi R_H^2(e-p)} \ , \phantom{spacing} 
   v_H = \frac{1+8\pi R_H^2p}{1-8\pi R_H^2e}  \ ,
 \label{eq_vH_BH}
\end{equation}
 (in this sub-section, $e$ and $p$ will always be evaluated at the 
horizon location and so we omit the subscript $H$ for them). For 
normal matter, not violating the null energy 
condition (NEC) $e+p\geq0$, the numerator of the expression for $\alpha$ 
is always positive and the sign of $\alpha$ (determining the signature of 
the horizon) is the same as that of the denominator (proportional to 
$-{\mathcal{L}_{-} \theta_{+}}$). The behaviour of this depends on the 
initial density profile of the configuration and on the equation of 
state, as we have seen from the simulation results presented here.

As we have already mentioned when discussing Figure \ref{2m_R}: 
for collapses where the $R=2M$ condition is first reached within bulk of the 
matter, under conditions which are non-singular, 
a single marginally trapped surface appears there with $\alpha=1$ and with 
the local value of the energy density being $e=1/2A_H$, independently of the 
pressure. As can be seen from Figure \ref{fig_BH(alpha-v_H)}, this is the only 
location in the diagram where a horizon with $e$ and $p$ non negative is 
neither ingoing or outgoing. The initial marginally trapped surface then 
separates into a pair of horizons, one outgoing and one ingoing, with 
$v_H\to\pm\infty$. The outgoing horizon, which is always spacelike, 
starts at the top of the upper branch of the hyperbola and then rolls 
down it, eventually reaching the stationary state corresponding to a null 
isolated horizon with $v_H=1$ when all of the collapsing matter has 
passed through it, so that $(e+p)=0$ at the horizon location. The ingoing 
horizon is also initially spacelike but starts at the bottom of the 
\emph{lower} branch of the hyperbola, and then rolls up it, possibly 
passing to the upper branch (the two branches are connected via 
$\alpha\rightarrow\pm\infty$) where it comes in from the left as a 
\emph{timelike} horizon and continues to roll upwards.

The change from spacelike to timelike corresponds to satisfying the 
condition $e-p=1/A_H$, for which the denominator of the left hand 
relation in Eq.(\ref{eq_vH_BH}) vanishes. Once again, we observe here a 
connection between fluid quantities (energy density and pressure) and a 
geometrical one (horizon area). Whether or not the change to timelike 
occurs seems to depend on the initial density profile of the 
configuration and on the equation of state. We have observed it happening 
for the constant-density initial models (with both $\gamma=4/3$ and 
$\gamma=5/3$) but not for the model starting from hydrostatic 
equilibrium, where the ingoing horizon always remained spacelike.

In each of our three cases, whether the ingoing horizon passes to the upper 
branch or not, it eventually reaches a maximum height on the plot and then 
rolls down the hyperbola again, ending on the lower branch, going back to 
$v_H \to - \infty$ and $\alpha=1$ from where it started. This is different 
from what happens with the the constant-density, zero-pressure 
Oppenheimer-Snyder collapse model, where the behaviour is characterised by 
constant values of $\alpha$ and $v_H$. However, as discussed above, 
that is a limiting case of the set of pressureless LTB models studied in 
\cite{Booth:2005ng} which have no density discontinuity at the surface and 
for many of which the horizon formation follows the standard picture 
described above (with $\alpha = 1$), even if other aspects of the evolution 
are significantly different from the situation with non-zero pressure. 
In one of the LTB models ($\sigma = 2$) where the Oppenheimer-Snyder limit 
is not approached in the centre, we see a second outgoing horizon emerging 
from the singularity formed at $r=0$ with $v_H \simeq 1$, rolling along the upper 
branch of the hyperbola up to $v_H\to+\infty$ where it meets and annihilates 
with the ingoing horizon which is rolling along the lower branch down to 
$v_H\to-\infty$. This is consistent with what we have seen in our runs near 
$\mu=0$, and shows the identical geometrical nature of the formation/annihilation 
processes for ingoing and outgoing horizons.

 We now give a coordinate-independent interpretation of the increase (or 
decrease) of the horizon area using Eq.(\ref{RWL2}), which gives
 \begin{equation}
 \mathcal{L}_{t^a}R_H  = \left.\frac{dR}{d\tau}\right\vert_H = U_H(1-v_H) \ ,
 \label{eq_areal_velocity_BH}
\end{equation}
 with $U$ always being negative because the matter is collapsing. Thus 
the area $A_H$ increases along the path of the horizon if $v_H>1$, and 
decreases if $v_H<1$. For a classical fluid, energy and pressure are both 
positive and the NEC is not violated. Under these circumstances the 
outgoing horizon is spacelike, (null just in the limit, as in the vacuum 
Schwarzschild solution) and $A_H$ is always increasing, while in general 
$A_H$ is always decreasing for an ingoing horizon. If the NEC is 
violated, with the pressure becoming negative, the outgoing horizon is 
allowed to roll down the upper branch of the hyperbola beyond $v_H=1$, 
becoming timelike with a decreasing $A_H$, signifying a shrinking 
surface. This happens in the presence of Hawking radiation, consistently 
with the fact that a timelike outgoing horizon ($0<v_H<1$), allows 
emission to go through the horizon \cite{Ellis:2013oka, 
Firouzjaee:2014zfa}. Note also the consistency of the horizon still being 
outgoing in the presence of Hawking radiation, remembering that $v_H$ is 
measured with respect to the collapsing matter.

In our simulations, we considered only classical matter {with} $e$ and $p$ 
positive which excludes one conceivable horizon configuration, i.e.\ the 
inner and outgoing horizon. Indeed, the latter simultaneously has 
$\mathcal{L}_{-}\theta_+\propto [(e-p)-1/A_H] > 0$ because it is inner, 
and $v_H>0$ because it is outgoing, and therefore from 
Eq.(\ref{eq_vH_BH}) we must have $p<-1/2A_H<0$. Hence for a classical 
fluid the outgoing horizon can only be outer. Whenever the NEC is 
satisfied (as it is for a classical fluid) there is a strict one-to-one 
correspondence between spacelike and outer on the one hand,
and timelike and inner on the other hand (see theorem 2 of 
\cite{Hayward:1993wb}, and \cite{Booth:2005ng}). Since the 
outgoing horizon can only be outer for classical matter, it must 
therefore be spacelike whereas the ingoing horizon can sometimes be 
spacelike (outer) and sometimes timelike (inner). Having this classically 
excluded configuration (inner and outgoing) seems to confirm the 
possibility of using the relation between $\alpha$ and $v_H$ also within 
the context of quantum effects (such as Hawking evaporation, mentioned 
before) which indeed allow {having negative pressure} and/or violation of 
the energy conditions.

The intersection with the horizontal axis is another special point of 
the hyperbola with $\alpha=-1$ and $v_H=0$, corresponding to the 
geometrical relation $p=-1/2A_H$, which for a homogeneous and isotropic 
fluid corresponds to $(e+3p)=0$. In this case the value of $A_H$ is 
independent of the energy density and depends only on the pressure, a 
symmetric relation to the one at the point of horizon formation ($\alpha=1$ 
and $v_H=\pm\infty$, with $e=1/2A_H$ independently of the value of the 
pressure). This point of the hyperbola represents the situation for a 
horizon which is neither outgoing nor ingoing, but is instead comoving with 
the collapsing matter. Rolling down the hyperbola, an outgoing horizon 
passing through this point would then become ingoing, while an ingoing 
horizon rolling up along the hyperbola would instead become outgoing. This 
corresponds to a turnaround of the horizon with respect to the matter.  
The location of the transition point for the ingoing horizon where it goes 
from positive values of the pressure to negative ones, depends on the 
particular value of the energy density. If this value is always smaller 
than or equal to the value for the homogeneous regime ($e=3/8\pi R^2$), as 
seen in Figure \ref{R2e_mu} for our simulations, the pressure will become 
negative for values of $v_H$ limited by the Oppenheimer-Snyder value (i.e.  
$v_H\leq-1/2$). Instead, if the energy density is reaching values larger 
than the homogeneous regime, then the pressure will change sign between the 
Oppenheimer-Snyder value and the point where the horizon is comoving with 
the matter (i.e $-1/2<v_H<0$). In Figure \ref{fig_BH(alpha-v_H)} the region 
of possible negative pressure has been highlighted in grey, degrading where 
the change of sign would occur.

If an ingoing horizon becomes outgoing with a turnaround and keeps rolling 
up to $v_H=1$ and $\alpha=0$, then from Eq.(\ref{eq_areal_velocity_BH}) 
$A_H$ will start increasing, signifying that the horizon bounces.  Such a 
bounce has been proposed as an alternative to the classical result of black 
hole singularity formation 
\cite{Frolov:1979,Roman:1983zza,Rovelli:2014cta,Barcelo:2014cla}. Although 
these scenarios are quite speculative, and we are not trying to draw any 
conclusion regarding them here, we think that it is worth pointing out that 
they could in principle be included within this sort of discussion of 
$\alpha$ and $v_H$. In future, we plan to investigate some of these 
possibilities using an equation of state including quantum phenomenology 
which allows for violation of the energy conditions.

\section{Summary and Conclusions}
\label{section_conclusion}
 In the present paper, we have brought together the 
Misner-Sharp-Hernandez hydrodynamical formalism for calculations in 
spherical symmetry and the geometrical formalism normally used for 
discussing trapping horizons.  We have given our motivations and 
explained why quasi-local horizons are useful in dynamical situations 
such as collapse to form black holes. By relating the expansion of the 
null geodesic congruence $\theta$ to the hydrodynamical parameters $U$ 
and $\Gamma$, we have confirmed that the horizons which appear in the 
Misner-Sharp-Hernandez formalism are precisely trapping horizons 
(we have used this term for 3D hypersurfaces as well as 2D surfaces here) 
and can be defined equivalently as loci where $U\pm\Gamma$ vanishes or 
where $\theta_{\pm}=0$, both corresponding to $R=2M$.

We then used this unified framework to study trapping horizons, focusing 
on the geometrical $\alpha$ (the sign of which gives the causal nature of 
the horizon), and the hydrodynamical $v_H$, which is the three velocity 
of the horizon with respect to the collapsing/expanding fluid. For a 
perfect fluid medium, each can be expressed as a simple algebraic 
function of the energy density and pressure of the fluid at the location 
of the horizon, and they can also be expressed as simple functions of 
each other.

We applied this formalism and notation to the study of black hole 
formation, presenting results from numerical simulations of spherically 
symmetric stellar collapse similar to those made in 1966 by May \& White 
\cite{May:1966zz} but focusing now on the behaviour of the trapping 
horizons. Following the line of the previous work, we investigated three 
sample cases, all including pressure effects using a polytropic equation of 
state. Two started from constant density but with different values of the 
adiabatic index $\gamma$ ($5/3$ and $4/3$), while the third, more 
realistic, case started from a configuration in hydrostatic equilibrium 
which was then made unstable to collapse by reducing the parameter $K$. We 
note that, in many respects, our results are rather different from those 
given by the analytic Oppenheimer-Snyder solution \cite{Oppenheimer:1939ue} 
for pressureless collapse from constant-density initial conditions. More 
elaborate studies of pressureless collapse, using LTB models with 
non-uniform density \cite{Booth:2005ng} show greater similarity with the 
present work but there are still significant differences and, in general, 
one should treat with caution the use of calculations with 
pressureless matter as a guide for realistic physical situations in the 
present context.

Our simulations all show the formation of a marginally trapped surface 
within the collapsing matter with $\alpha=1$ which then separates into two 
parts, one moving outwards ($v_H>0$) and the other moving inwards 
($v_H<0$) with respect to the matter. Both of these horizons form spacelike 
with $v_H \to \pm \infty$. 
For any classical fluid not violating the NEC, it is known that the 
outgoing horizon (which is an outer horizon according to Hayward's 
terminology) must remain spacelike while passing through the collapsing 
matter and this is indeed seen in the simulations with the horizon 
eventually becoming null when all of the collapsing matter has passed 
through it. At the final stage, it is an isolated horizon in vacuum, 
equivalent to the event horizon of the Schwarzschild solution. For the two 
cases starting from constant density, the ingoing horizon starts spacelike, 
subsequently becomes timelike, and then goes back to being spacelike again 
at the end; for the case starting from hydrostatic equilibrium, it always 
remains spacelike. In all cases, the ingoing horizon continues to move 
towards $\mu=0$, finally shrinking away there with $\alpha \to 1$ and $v_H 
\to -\infty$, the same conditions as when it was formed. 
Plotting the functional relationship between $\alpha$ and $v_H$ gives 
a rectangular hyperbola, and we found that this gives a convenient way of 
exhibiting the horizon evolution, leading to additional insights. The roles 
of differing initial configurations and differing equations of state 
require further investigation with a more systematic analysis, and we are 
now embarking on that. \\

\ack In the process of preparing this paper, we have benefited from 
discussions with a number of colleagues; in particular, we are very 
grateful to George Ellis, \mbox{Valerio} Faraoni, Javad 
\mbox{Firouzjaee}, Luciano Rezzolla, Stefano Liberati, Pierre 
Bin\'{e}truy, Fr\'{e}d\'{e}ric Lamy, Carlo Rovelli, Francesca 
\mbox{Vidotto}, Pier \mbox{Stefano} Corasaniti, and \'{E}ric Gourgoulhon. 
I.M. and J.M. thank the PCCP group of APC for hosting them during the 
course of the collaboration, and I.M. also thanks the Astrophysics 
sub-department of the University of Oxford for their hospitality. The 
research leading to these results has received funding from the European 
Research Council under the European Community's Seventh Framework Program 
(FP7/2007-2013) StG-EDECS (Grant Agreement No. 279954) and from the ERC 
Advanced Grant 339169 ``Self-Completion''.

\appendix
\section{Misner-Sharp-Hernandez equations}
\label{Appendix_MSH}
 Here we present the Misner-Sharp-Hernandez equations in the 
composite form based on \cite{Misner:1964je,Misner:1966hm,May:1966zz} 
which we have used for the work reported in this paper. Consider the 
``cosmic time'' metric given by Eq.(\ref{eq_metric_MS}) with the definitions of 
$U$, $\Gamma$ and $M$ given in Eqs.(\ref{U_def}), (\ref{Gamma_def}), 
(\ref{eq_MS_energy}) and a perfect fluid with stress energy tensor 
$T^{a}_{\phantom{a}b} =\textrm{diag}(-e,p,p,p)$ (consistent with 
Eq.(\ref{eq_stress})), where $e$ is the energy density and $p$ is the 
pressure. Then the Misner-Sharp-Hernandez hydrodynamic equations 
obtained from the Einstein equations and the conservation of the stress 
energy tensor are:
 \begin{eqnarray}
& D_t U = - \frac{\Gamma}{e+p} D_r p - \frac{M}{R^2} - 4\pi Rp \,, 
\label{Euler_eq} \\
& D_t \rho= -\frac{\rho}{\Gamma R^2}D_r(R^2U) \,,\label{D_trho} \\
& D_t e = \frac{e+p}{\rho} D_t \rho \,, \label{en_eq} \\
& D_r a = -\frac{a}{e+p}D_r p \,, \label{lapse_eq} \\
& D_r M = 4\pi R^2\Gamma e \,, \label{D_rM}
\end{eqnarray}
 where $\rho$ in Eqs.(\ref{D_trho}) and (\ref{en_eq}) is the rest mass 
density (or the compression factor for a fluid of particles without rest 
mass). These form the basic set, together with the constraint equation 
given already by Eq.(\ref{eq_MS_mass}),
 \begin{equation}
\Gamma^2 = 1 + U^2 - \frac{2M}{R} \,.
\label{constraint_eq}
\end{equation}
Two other useful expressions coming from the Einstein equations are
 \begin{eqnarray}
& D_t \Gamma = -\frac{U}{e+p} D_r p \,, \label{D_tGamma} \\
& D_t M = - 4\pi R^2Up \,. \label{D_tM}
 \end{eqnarray}
 In order to derive the expression for $\alpha$ given by Eq.(\ref{alpha}) 
we need to make some manipulation of these equations. Combining 
Eqs.(\ref{Euler_eq}) and (\ref{D_tGamma}) gives
 \begin{equation}
D_tU \pm D_t\Gamma = - \frac{\Gamma \pm U}{e+p}D_rp 
 - \frac{M}{R^2} - 4\pi Rp \,.
\label{D_tU-D_tGamma}
\end{equation}
 Differentiating $\Gamma$ with respect to $r$ in Eq.(\ref{constraint_eq}) 
gives
 \begin{equation}
D_r\Gamma = \frac{U}{\Gamma}D_rU + \frac{M}{R^2} - \frac{1}{\Gamma R}D_rM \,,
\end{equation}
 that combined with Eq.(\ref{D_rM}) allows one to write
\begin{equation} 
D_rU \pm D_r\Gamma = \frac{\Gamma \pm U}{\Gamma}D_rU \pm \frac{M}{R^2} 
\mp 4\pi R e \,.
\label{D_rU-D_rGamma}
\end{equation}
 Using now expressions (\ref{D_tU-D_tGamma}) and (\ref{D_rU-D_rGamma}) 
appearing in Eqs.(\ref{alpha_BH}) and (\ref{alpha_CH}) with the 
corresponding conditions for the black hole horizons ($\Gamma=-U$) and 
for the cosmological horizon in an expanding universe ($\Gamma=U$), 
we obtain
 \begin{equation}
 \alpha=\left.\frac{4\pi R^2(e+p)}{1-4\pi R^2(e-p)} \right\vert_H 
 \end{equation}
in both cases, which is Eq.(\ref{alpha}).

\section{Equation of state}
\label{Appendix_eqstate}
 In order to solve the set of equations presented in the previous 
Appendix we need to supply an equation of state specifying the relation 
between the pressure and the different components of the energy density. 
For a simple ideal particle gas, we have that
 \begin{equation}
 p = (\gamma-1)\rho\epsilon \,,
 \end{equation}
 where $\epsilon$ is the specific internal energy, related to the velocity 
dispersion (temperature) of the fluid particles and $\gamma$ is the 
adiabatic index. The total energy density $e$ is the sum of the rest mass 
density and the internal energy density:
 \begin{equation}
e = \rho(1+\epsilon) \,.
\label{eq_state}
\end{equation}
 Putting these equations into the energy equation (\ref{en_eq}), one 
gets the standard polytropic form used for stellar models
 \begin{equation}
p=K\rho^{\gamma}\,,
\label{polytropic}
\end{equation}
 where $K$ is a constant of integration (varying with the specific entropy if 
 the process is not adiabatic), and $\gamma$ is the adiabatic index depending 
 on the type of the matter.
 
In general, if $\gamma\neq1$, Eq.(\ref{eq_state}) can be written as
 \begin{equation}
e = \rho + \frac{p}{\gamma-1}
\end{equation}
 showing that, when the contribution of the rest mass of the particles to 
the total energy density is negligible ($e\gg\rho$, $\epsilon\gg1$) we get the 
standard (one-parameter) equation of state used for a cosmological fluid
 \begin{equation}
p=we
\label{cosmo_eqstate}
\end{equation}
 setting $w=\gamma-1$. A pressureless fluid ($w=0$) corresponds to the 
case where the specific internal energy $\epsilon$ is effectively zero.

In the case of Eq.(\ref{cosmo_eqstate}) the equation of state has a 
constant ratio of pressure over energy density given by $w$, while in the 
polytropic case given by Eq.(\ref{polytropic}) this ratio is varying with 
the density, increasing during the collapse. For an ideal gas in general 
we have
\begin{equation}
\frac{p}{e} = \frac{\epsilon}{1+\epsilon}(\gamma-1)\,,
\end{equation}
 varying from $\epsilon (\gamma-1)$ when $\epsilon\ll1$ to the limit of 
$w$ when $\epsilon\gg1$.

\end{document}